\documentclass[%
preprint,
superscriptaddress,
amsmath,amssymb,
aps,
]{revtex4-2}

\usepackage{graphicx}% Include figure files
\usepackage{dcolumn}% Align table columns on decimal point
\usepackage{bm}% bold math
\usepackage{titlesec}
\usepackage[caption=false]{subfig} %subfloat
\usepackage[colorlinks=true]{hyperref} %color in references to fig,bib,etc
\hypersetup{colorlinks,
	citecolor=blue
}
\usepackage{cleveref}
\usepackage{booktabs} 
\usepackage{array}
\usepackage{tikz}
\usepackage{ulem}
\usepackage{slashed}
\usepackage{ulem}

\usetikzlibrary{calc,shapes.misc,shapes.geometric}
\newcommand\Mark[2][]{%
\tikz[baseline=(a.base)]{
\node[inner sep=0pt,outer sep=0pt](a){\phantom{#2}};  %% just to be defensive
\node[draw,black,thick,inner sep=2pt, rectangle,text=black,overlay,#1]  {#2};%
}
}
\newcommand{\overbar}[1]{\mkern 1.5mu\overline{\mkern-1.5mu#1\mkern-1.5mu}\mkern 1.5mu}
\Crefname{figure}{Fig.}{Figs.}

\newcommand{\eq}[1]{Eq.~\eqref{#1}}

\begin{document}

\title{Final state interactions in semi-inclusive neutrino-nucleus scattering: Application to T2K and MINER$\nu$A experiments}

%\title{Comparison of SuSAv2-MEC implementation in GENIE and an unfactorized relativistic approach to semi-inclusive neutrino-nucleus interactions with T2K and MINER$\nu$A measurements \\\vspace{0.25cm} \color{red}or\color{black} \\\vspace{0.25cm}
%Comparison of SuSAv2-MEC implementation in GENIE with an unfactorized relativistic approach to semi-inclusive neutrino-nucleus interactions
% \\\vspace{0.25cm}
%{\ttblue M:  I think  the first title would be better, if the next step is to do the same for argon. However, I have some obejctions: 1) it is very long, 2) it contains a lot of acronyms, 3) it does not have information on FSI, which are the focus of this paper as a follow-up of the previous "Theoretical description of semi-inclusive T2K, Miner?A and MicroBooNE neutrino-nucleus data in the relativistic plane wave impulse approximation", JMF,  Phys.Rev.D 104 (2021) 7, 073008 and 4) this is a theory paper, I would put more emphasis on the theoretical aspects than on the comparison with data.  What about "Final state interactions in semi-inclusive neutrino-nucleus scattering"?}}

\author{J. M. Franco-Patino}
\affiliation{
Departamento de Física atómica, molecular y nuclear, Universidad de Sevilla, 41080 Sevilla, Spain
}
\affiliation{
	Dipartimento di Fisica, Università di Torino, Via P. Giuria 1, 10125 Torino, Italy
}
\affiliation{
	INFN, Sezione di Torino,  Italy
}

\author{R. González-Jiménez}
\affiliation{
	Grupo de Física Nuclear, Departamento de Estructura de la Materia, Física Térmica y Electrónica and IPARCOS, Facultad de Ciencias Físicas, Universidad Complutense de Madrid, CEI Moncloa, Madrid 28040, Spain
}

\author{S. Dolan }
\affiliation{CERN, European Organization for Nuclear Research, Geneva, Switzerland }

\author{M. B. Barbaro}
\affiliation{
	Dipartimento di Fisica, Università di Torino, Via P. Giuria 1, 10125 Torino, Italy
}
\affiliation{
	INFN, Sezione di Torino,  Italy
}
\affiliation{IPSA Paris, 63 boulevard de Brandebourg,
94200 Ivry-sur-Seine, France}

\author{J. A. Caballero}
\affiliation{
	Departamento de Física atómica, molecular y nuclear, Universidad de Sevilla, 41080 Sevilla, Spain
}
\affiliation{
	Instituto de Física Teórica y Computacional Carlos I, Granada 18071, Spain
}

\author{G. D. Megias}
\affiliation{
Departamento de Física atómica, molecular y nuclear, Universidad de Sevilla, 41080 Sevilla, Spain
}
\affiliation{
	Research Center for Cosmic Neutrinos, Institute for Cosmic Ray Research, University of Tokyo, Kashiwa, Chiba 277-8582, Japan
}

\author{J. M. Udias}
\affiliation{
	Grupo de Física Nuclear, Departamento de Estructura de la Materia, Física Térmica y Electrónica and IPARCOS, Facultad de Ciencias Físicas, Universidad Complutense de Madrid, CEI Moncloa, Madrid 28040, Spain
}

\date{\today}

\begin{abstract}
	We present a complete comparison of semi-inclusive $\nu_\mu$-$^{12}$C cross-section measurements by T2K and MINER$\nu$A collaborations with the predictions from the SuSAv2-MEC model implemented in the neutrino-nucleus event generator GENIE and an unfactorized approach based on the relativistic distorted wave impulse approximation (RDWIA). Results, that include cross sections as function of the final muon and proton kinematics and correlations between both, show that the agreement with data obtained by the RDWIA approach, that accounts for final-state interactions, matches or improves GENIE-SuSAv2 predictions for very forward angles where scaling violations are relevant.
\end{abstract}

\maketitle

\section{\label{sec:1}Introduction}
	Accelerator-based neutrino oscillation experiments such as T2K~\cite{PhysRevD.98.032003,nature,Abe:2011ks}, MINER$\nu$A~\cite{PhysRevLett.121.022504,MINERvA:2021csy,MINERvA:2013zvz}, DUNE~\cite{DUNE:2020ypp} and Hyper-K~\cite{10.1093/ptep/ptv061} offer an unprecedented opportunity to explore fundamental physics, such as the charge-parity (CP) violation in the lepton sector, neutrino mass hierarchy and physics beyond the standard model~\cite{nature,review}, although their success relies on understanding neutrino-nucleus interactions in the energy range of a few GeV. In these experiments the incoming neutrino energy distribution is a broad function, thus the exact energy of the interacting neutrino is unknown. For a reliable analysis of the interaction of a neutrino with a nucleus, all the possible reaction channels that contribute to the experimental signal need to be taken into account. This difficulty is less severe in electron scattering experiments, where the incoming beam of electrons has a well-defined energy and the kinematics can be selected to separate different reaction mechanisms in  the nuclear response.
  
  	The T2K experiment reconstructs the neutrino energy from the measured lepton kinematics. The main analysis samples aim to identify charged-current meson-less neutrino interactions (CC0$\pi$) and then reconstruct the neutrino energy assuming the interactions are charged-current quasielastic (CCQE) scatters off a single nucleon at rest with some fixed binding energy~\cite{ T2K:2021xwb}. T2K then uses the NEUT neutrino-nucleus interaction simulation~\cite{Hayato:2021heg} in order to estimate the significant reconstruction biases due to nuclear effects (beyond a fixed binding energy) and from non-CCQE contributions to the CC0$\pi$ sample, such as those in which a neutrino scatters off a bound state of two nucleons (two-particle-two-hole or 2p2h excitations) or those that produce a pion which is absorbed inside the nuclear medium~\cite{PhysRevD.85.113008, Alvarez_Ruso_2014}. In order to reliably infer the oscillated neutrino energy spectra at the far detector (which is crucial for characterizing neutrino oscillations), it is therefore essential that the modelling of nuclear effects, 2p2h and pion absorption are under control and that their plausible variation is covered by a robust estimation of theoretical uncertainties.
  	
	Measurements of outgoing lepton kinematics in CC0$\pi$ events are very important for experiments like T2K and Super-Kamiokande~\cite{Abe:2011ks,Fukuda:2002uc} where most of the information about the oscillation signal comes from detection of the final-state muons only. However, they do not allow to discriminate between different nuclear models and are not sufficient to put constraints on the amount of two-body current contributions. This is why there is a growing interest in measurements of more exclusive processes, for instance the detection in coincidence of a muon and an ejected proton in the final state. The interpretation of such reactions, usually called semi-inclusive reactions~\cite{PhysRevD.90.013014, PhysRevC.100.044620, PhysRevC.102.064626, Ershova:2022jah}, is challenging as it requires realistic descriptions of the initial nuclear state and a good control of proton final-state interactions (FSI) in Monte Carlo event generators. From semi-inclusive neutrino-nucleus events one could reconstruct the energy of the incoming neutrino in a region where the missing energy is well-defined, which is the case of CCQE scattering where the neutrino scatters off a single bound nucleon and the missing energy is of the order of the binding energy of the nucleon. In this context, a recent study~\cite{PhysRevD.105.032010} has shown that a neutrino energy estimator depending on the muon but also on the final proton kinematics, although neglecting nuclear removal energy and the loss of energy due to nucleon FSI, improves the reconstructed energy resolution and the sensitivity to possible bias in the removal energy estimation.
	
	The implementation of neutrino interaction models in neutrino event generators requires a fast method of calculating the differential cross section given some set of outgoing particle kinematics. In the best scenario, the full exclusive cross section should be available as function of all the variables that define the final state, which are five in the specific case of one proton knockout reaction. However, there are very few unfactorized microscopic models which take into account FSI and that are suitable to compare to inclusive or exclusive cross sections measurements. None of these unfactorized models are currently implemented in any neutrino event generator which, instead, usually rely on factorization approaches, stemming from the plane-wave impulse approximation (PWIA). One of these unfactorized microscopic models, extensively applied in the past to describe exclusive electron scattering $\left(e,e'p\right)$ cross-section measurements within a relativistic and fully quantum approach based on the relativistic distorted-wave impulse approximation (RDWIA)~\cite{PhysRevC.48.2731, PhysRevC.51.3246, PhysRevC.64.024614,Amaro21}, uses a relativistic optical potential (ROP) to include FSI. In this case the outgoing nucleon is described by a scattering wave solution of the Dirac equation with this ROP, that includes real and imaginary terms that are fitted to reproduce elastic proton-nuclei scattering data. Another possibility is to characterize the final nucleon as a scattering solution of a relativistic mean field (RMF) potential parameterized to reproduce properties of nuclei.
	
	In contrast with the microscopic and unfactorized models like RDWIA, which incorporate in the modelling both the lepton-boson and the boson-nucleus vertex in some detail, and thus can be compared to semi-inclusive observables,  there are other models, like the SuSAv2 model~\cite{PhysRevC.90.035501, PhysRevD.94.013012, PhysRevC.99.042501}, that are aimed to describe inclusive cross sections, that is, only as function of the final lepton kinematics and thus cannot make predictions on both leptons and hadrons in the final state. In spite of this, by taking advantage of a factorization approach, some neutrino event generators like GENIE~\cite{ANDREOPOULOS201087, andreopoulos2015genie} can make predictions about the lepton and also the outgoing nucleon kinematics from these inclusive models~\cite{PhysRevD.101.033003, dolan2021implementation}. In GENIE, exploiting a factorization approximation, for a given event the inclusive models provide the lepton variables, while the kinematics of the ejected proton can be determined by selecting an initial nucleon from a local Fermi gas distribution and then applying momentum and energy conservation at the vertex. This procedure implies that the initial nuclear state, which takes part in the event, is decoupled from the leptonic vertex. This way, while the behavior of the cross section against the muon kinematics may be described correctly, there is no guarantee whatsoever that the correlations between final muons and protons for a given event are preserved. Moreover, this approach could give inconsistent results when the nuclear model used to generate the outgoing nucleon is different from the nuclear model used in the inclusive cross section, as is the case in the current SuSAv2 implementation in GENIE~\cite{PhysRevD.101.033003}. Furthermore, the results from this approach rely strongly on the semi-classical description of FSI commonly used in neutrino event generators~\cite{doi:10.1063/1.3661588,PhysRevC.86.015505}, which have been shown to be unable to produce microscopic FSI predictions at low outgoing nucleon momenta~\cite{nikolakopoulos2022benchmarking}.
	
	In this paper we will extend previous analyses of T2K~\cite{PhysRevD.98.032003} and MINER$\nu$A~\cite{PhysRevLett.121.022504, PhysRevD.101.092001} $\nu_\mu-^{12}$C semi-inclusive CC0$\pi$ cross-section measurements with one muon and at least one proton in the final state (denoted CC0$\pi$Np) in PWIA~\cite{PhysRevD.104.073008} to include FSI within a fully relativistic, quantum mechanical and unfactorized approach using both a ROP fit to elastic proton-nucleus scattering data, and a modified version of the RMF potential~\cite{PhysRevC.100.045501, PhysRevC.101.015503, Gonzalez-Jimenez:2021ohu}, to describe the proton in the final state. We will compare our microscopic results with the estimations from the inclusive SuSAv2-MEC model implemented in GENIE event generator~\cite{PhysRevD.101.033003} and test the validity of the approximations made by the event generators to obtain hadron kinematics using as starting point such an inclusive model. Due to the fact that we do not have semi-inclusive 2p2h MEC and pion absorption models, we will add the SuSAv2-2p2h MEC~\cite{RUIZSIMO2016124, PhysRevD.91.073004, PhysRevD.94.093004} and pion absorption contribution calculated with GENIE to our QE results for a full comparison with the available cross-section measurements. Estimates presented in \cite{PhysRevD.98.032003, PhysRevD.101.033003} suggest that the main non-QE contribution to the T2K semi-inclusive cross sections as function of the muon and proton kinematics is the 2p2h MEC channel. The next non-QE contribution comes from production of pions that are absorbed inside the nucleus. The latter is small for T2K, although this is not the case for MINER$\nu$A due to the higher energy of the neutrinos.
	
	We summarize in Sec.~\ref{sec:2} the general formalism of semi-inclusive neutrino-nucleus reactions as well as the description of the initial state and the final-state models considered in this work. In Sec.~\ref{sec:3} we briefly address the implementation of SuSAv2 model in the neutrino event generator GENIE and the approximations used to allow inclusive models to describe semi-inclusive reactions. In Sec.~\ref{sec:piabs} we give a short description of the pion absorption model of GENIE. Finally, in Sec.~\ref{sec:4} we present and compare the results of both approaches with T2K and MINER$\nu$A semi-inclusive CC0$\pi$Np cross-section measurements.
	
\section{\label{sec:2}Semi-inclusive neutrino-nucleus reactions within the impulse approximation}

\begin{figure}[!htbp]
		\centering
		\includegraphics[width=0.45\textwidth]{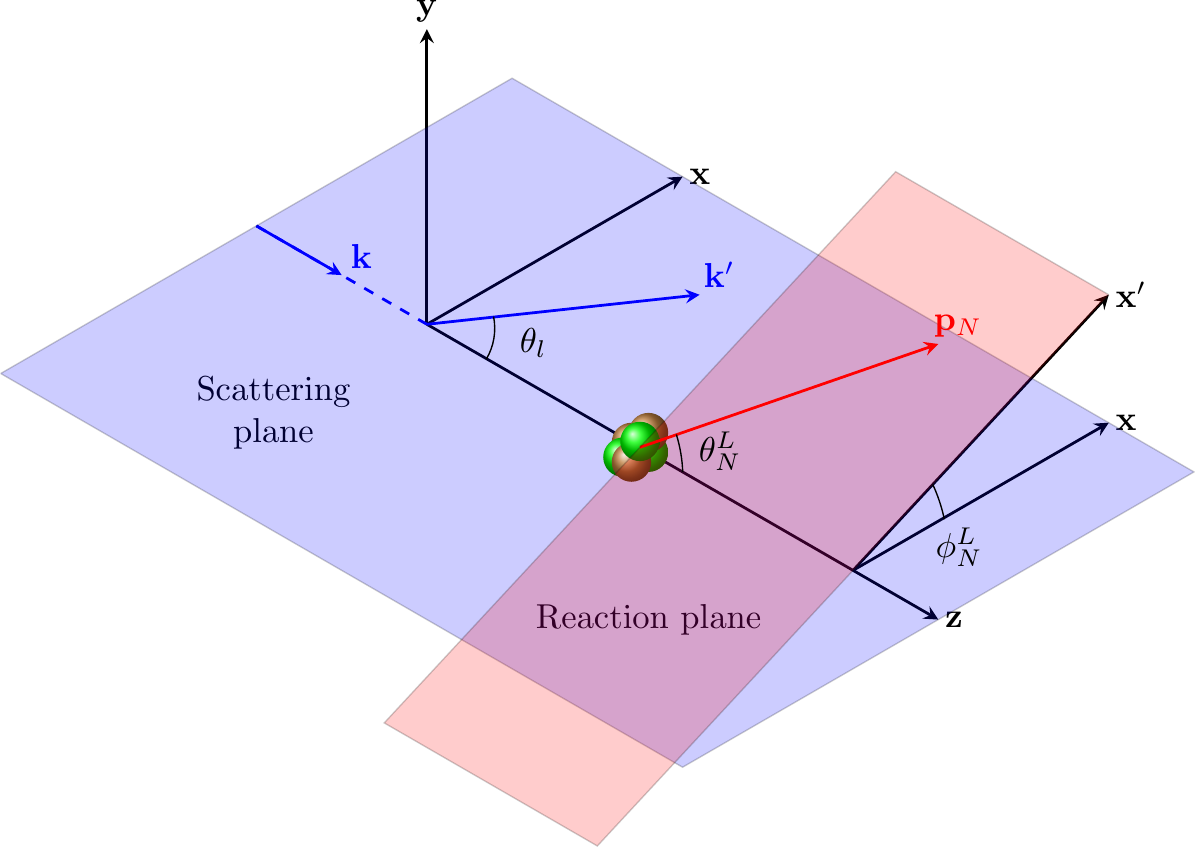}
		\caption{\label{fig:rs} Definition of the laboratory frame where the initial nucleus is considered at rest and the neutrino direction $\hat{k}$ is chosen to be the $z$-axis. The final lepton is defined by the momentum $k'$ and the scattering angle $\theta_l$, which is the angle with respect to the initial neutrino direction. The final nucleon is characterized by a momentum $p_N$ and two angles $\theta_N^L$ and $\phi_N^L$. The incoming neutrino ($\mathbf{k}$) and the final lepton ($\mathbf{k}'$) momenta are contained in the scattering plane, while the reaction plane contains the incoming neutrino and the ejected nucleon ($\mathbf{p_N}$).}
	\end{figure}

	In what follows, we assume that after the interaction of an incoming neutrino of momentum $\mathbf{k}$ with a nucleus $A$ we detect in the final state a lepton and a nucleon in coincidence, having momenta $\mathbf{k'}$ and $\mathbf{p_N}$, respectively. We consider that no other particles are detected in the final state, although they might be present depending on the kinematics. The kinematics of the outgoing lepton and nucleon for semi-inclusive CC$0\pi$ events is characterized by a set of six independent variables $\left(k',\theta_l,\phi_l,p_N,\theta_N^L,\phi_N^L\right)$ defined in Fig.~\ref{fig:rs} together with the laboratory frame where we will work, in which the cross section does not depend on $\phi_l$. We consider that the incoming neutrinos are distributed according to an energy distribution or flux $\Phi(k)$ and that the impulse approximation (IA) is valid, {\it i.e.}, the incoming neutrino interacts only with one of the bound nucleons exchanging a charged $W$ boson, and being knocked out of the nucleus, this is the nucleon detected. Then, the flux-averaged semi-inclusive neutrino-nucleus cross section is \cite{Gonzalez-Jimenez:2021ohu, nikolakopoulos2022benchmarking}
	\begin{widetext}
		\begin{align}\label{semi-inclusive}
		\left <\frac{d\sigma}{dk'd\Omega_{k'}dp_{N}d\Omega^{L}_{N}}\right >&=\frac{G_F^2\cos^2{\theta_c}k'^2p_N^2}{64\pi^5}\int dk\frac{W_B}{E_Bf_{\text{rec}}}L_{\mu\nu}H^{\mu\nu} \, \Phi(k) ,
		\end{align}
	\end{widetext}
	where $\Omega_{k'}$ and $\Omega_{N}^L$ are, respectively, the solid angles of the final lepton and the ejected proton, the residual nucleus $B$ can be left in an excited state with invariant mass $W_B$ and total energy $E_B$ , $L_{\mu\nu}$ and $H^{\mu\nu}$ are the leptonic and hadronic tensors, and $f_{\text{rec}}$ is the recoil factor given by
	\begin{align}
		f_{\text{rec}} = \left| 1- \frac{\mathbf{p_m}\cdot\hat{z}}{E_B}\right|
	\end{align}
	with $\mathbf{p_m} = \mathbf{q}-\mathbf{p_N}$ the missing momentum and $\mathbf{q} = \mathbf{k} - \mathbf{k'}$ the  transferred momentum. Note that the integral over the neutrino momentum on \eq{semi-inclusive} is equivalent to the integral over the so-called missing energy $E_m$ because they are related through the energy-momentum conservation:
	\begin{align}
		E_m = W_B + m_N - M_A = k - E_l + m_N - E_N - T_B
	\end{align}
	with $E_l$ the energy of the final lepton, $T_B$ the kinetic energy of the recoiling system, and $E_N$ and $m_N$ the total energy and mass of the final nucleon, respectively. 	 
	 
	All information about the nuclear structure and FSI effects is contained inside the hadronic tensor which is constructed as the bilinear product of the matrix elements of the nuclear current operator $\hat{J^\mu}$ between the initial nuclear state $\left|A\right>$ and the final hadronic state $\left|B,p_N\right>$, composed of the ejected nucleon and the undetected final nucleus,
	\begin{align}\label{hadronic_tensor_unfactorized}
		H^{\mu\nu} = \overline{\sum_{if}} J^\mu {J^{\nu}}^{\dagger} =\overline{\sum_{if}} \left<B,p_N\left|\hat{J}^\mu\right|A\right>\left<B,p_N\left|\hat{J}^\nu\right|A\right>^*,
	\end{align}
	where $\overline{\sum}_{if}$ corresponds to the appropriate average over initial states and sum over final states as discussed below. 
	Taking into account that we will describe the initial state as a product of RMF single-particle states labeled with a quantum number $\kappa$, we introduce the hadron tensor for each shell $\kappa$, given by
	\begin{align}\label{eq:hadronic_tensor}
%		H_\kappa^{\mu\nu} = \rho_\kappa\left(E_m\right)\sum_{m_j, s_N}J_{\kappa, m_j,    s_N}^{\mu}J_{\kappa, m_j, s_N}^{\nu\dagger},
		H_\kappa^{\mu\nu} = \rho_\kappa\left(E_m\right)\sum_{m_j, s_N}J_{\kappa, m_j, s_N}^{\mu}J_{\kappa, m_j, s_N}^{\nu*}
	\end{align}
	%{\ttblue [M: again, $\dagger\to *$]} 
	with $\rho_\kappa\left(E_m\right)$ the missing energy density and
	\begin{widetext}
		\begin{align}\label{eq:current}
			J_{\kappa, m_j, s_N}^\mu = \int d\mathbf{r}\, e^{i\mathbf{r}\cdot\mathbf{q}}\overbar{\Psi}^{s_N}\left(\mathbf{p_N},\mathbf{r}\right)\left(F_1\gamma^\mu +\frac{iF_2}{2m_N}\sigma^{\mu\nu}Q_\nu \right.
			 \left. + G_A\gamma^\mu\gamma^5 + \frac{G_P}{2m_N}Q^\mu\gamma^5\right) \Phi_\kappa^{m_j}\left(\mathbf{r}\right)\, ,
		\end{align}%left and right if widetext
	\end{widetext}
	where $m_j$ is the third component of the total angular momentum $j$ of the bound nucleon, $s_N$ the spin projection of the final nucleon and $Q^{\mu}$ the four-momentum transfer. The wave functions $\Psi^{s_N}$ and $\Phi_\kappa^{m_j}$ are four-dimensional spinors which describe, respectively, the scattered and bound nucleon and we have used the common CC2 expression for the one-body current operator~\cite{PhysRevC.102.064626}.

	In standard PWIA the differential cross section may factorize into an elementary cross section, describing lepton-nucleon scattering, and a spectral function describing the probability of finding a nucleon in the target nucleus with energy and momentum compatible with the kinematics of the process (see~\cite{PhysRevD.72.053005} for more details). This factorized result, although strictly valid only in PWIA, is useful for interpretation of the experimental data. The factorization approach makes it easy the use of sophisticated microscopic approaches to the spectral function to compare with the effective cross section derived from the analysis of data. Unfortunately, the simplicity of the factorized result is lost when distortion (that is, departure from plane waves) of either lepton and/or ejected nucleon wave functions are taken into account. Furthermore, factorization does not hold in the relativistic plane wave limit because of the role played by the negative energy components in the bound relativistic nucleon wave function~\cite{Caballero98a}.

\subsection{\label{subsec:2-1}Initial state description}

	The bound nucleons are described in our approach by a product of single-particle states $\Phi_\kappa^{m_j}\left(\mathbf{r}\right)$ obtained by solving the Dirac equation in coordinate space in presence of two RMF potentials $S(r)$ and $V(r)$ fitted to the nuclear ground state properties~\cite{WALECKA1974491,HOROWITZ1981503}.
	
\begin{figure}[!htbp]
		\centering
		\includegraphics[width=0.5\textwidth]{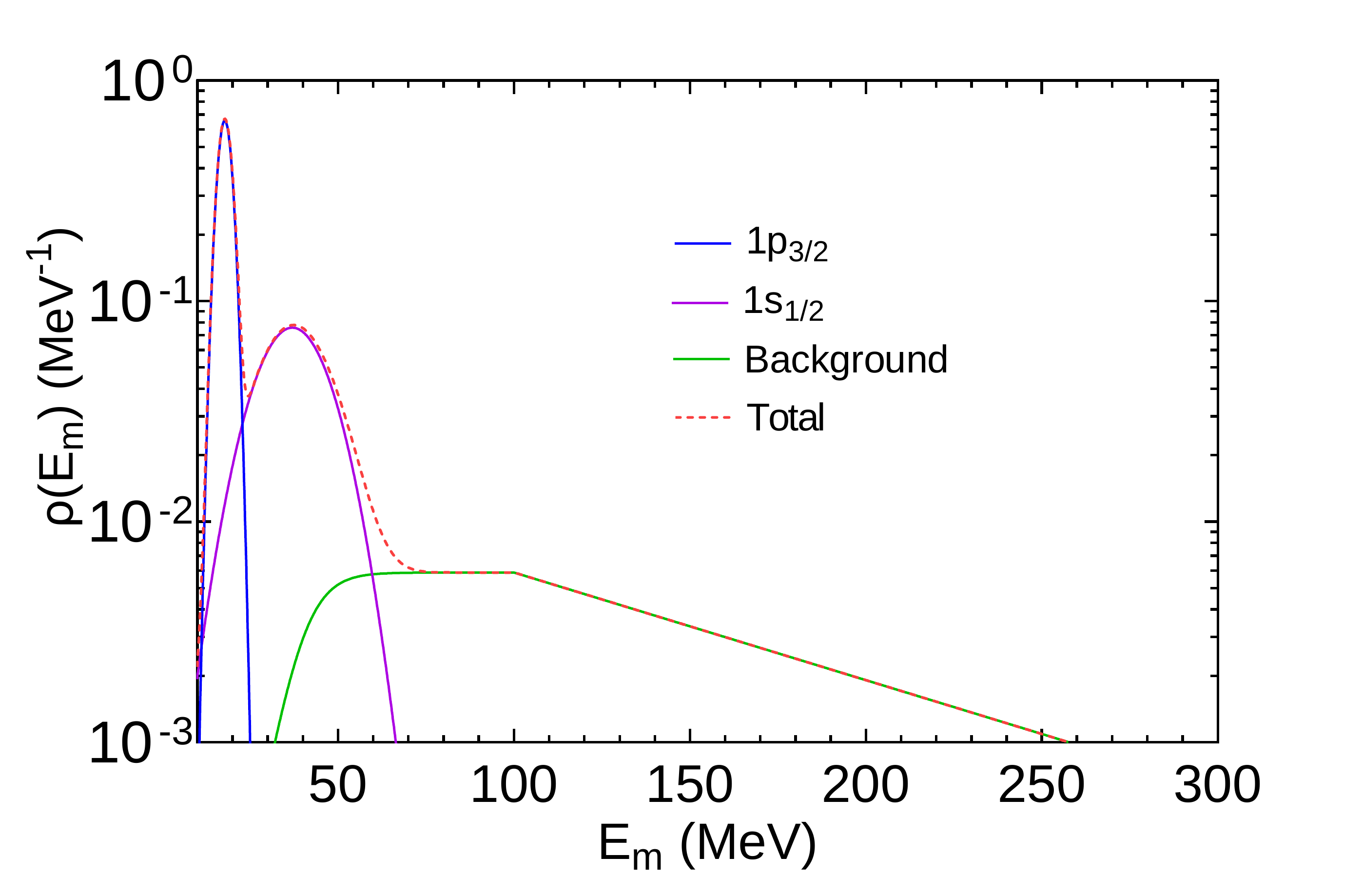}
		\caption{\label{fig:rho}Parameterization of $\rho_\kappa(E_m)$ for $^{12}$C by shells used in this work. The occupation numbers for the $1s_\frac{1}{2}$ and $1p_\frac{3}{2}$ shells are 1.9 and 3.3 respectively, with the remaining 0.8 nucleons associated to the background.}
	\end{figure}

	Effects beyond pure shell model approach can be introduced in the same fashion as~\cite{Gonzalez-Jimenez:2021ohu} for $^{16}$O. We will introduce a depletion of the occupation of the shell model states, and include high missing energy nucleons originating from correlations in the initial state. The missing energy density $\rho_\kappa(E_m)$ for $^{12}$C used in this work is shown in Fig.~\ref{fig:rho}. By using this method, we can include effects caused by long- and short-range correlations as seen in the spectral function formalism but without imposing factorization of the cross section. The parameters in the $\rho_\kappa(E_m)$ function used in this work were fitted to reproduce the missing energy profile that one gets from the Rome spectral function for $^{12}$C~\cite{BENHAR1994493, HOROWITZ1981503, PhysRevD.72.053005}. In this case, $\rho_\kappa(E_m)$ is composed by the contributions coming from the 1$s_{\frac{1}{2}}$ and the $1p_{\frac{3}{2}}$ shells, which are parameterized as Gaussian distributions as shown in Table~\ref{table: argon}, and from the background, which is considered as an additional s-shell parameterized as follows:
	
	\begin{equation}
		F\left(E_m\right) = a\:\textrm{exp}\left(-b \: E_m\right),	
	\end{equation}  
	if $E_m > 100$ MeV, and
	
	\begin{equation}
		F\left(E_m\right) = \frac{a\:\textrm{exp}\left(-100\:b\right)}{\textrm{exp}\left[-\left(E_m - c\right)/w\right] + 1},
	\end{equation}
	if $26 < E_m < 100$ MeV. The value of the parameters are $a = 0.031127\:\textrm{MeV}^{-1}$, $b = 0.011237\:\textrm{MeV}^{-1}$, $c = 40$ MeV and $w = 5$ MeV.
	\begingroup
	\setlength{\tabcolsep}{6.3pt}
	\begin{table}[!h]
		\centering
		\begin{tabular}{cccccccc}
			\toprule\toprule
			$\kappa$ & &$\mu_\kappa$ (MeV) & &$\sigma_\kappa$ (MeV) & &$n_\kappa $ \\\midrule
			$1s_{1/2}$ & &37.0 & &10.0 & &1.9\\\midrule
			$1p_{3/2}$ & &17.8 & &2.0 & &3.3\\\midrule					
			\bottomrule\bottomrule
		\end{tabular}
		\caption{\label{table: argon}Parameterization of the missing energy distributions for the two shells of $^{12}$C. The contribution to the missing energy density $\rho_\kappa\left(E_m\right)$ of each shell is given by $\rho_\kappa\left(E_m\right) = \frac{n_\kappa}{\sqrt{2\pi}\sigma_\kappa}\exp(-\left(\frac{E_m - \mu_\kappa}{2\sigma_\kappa}\right)^2)$, with $\mu_\kappa$ the mean value, $\sigma_\kappa$ the standard deviation and $n_\kappa$ the occupation number.}
	\end{table}
	\endgroup	
	
	In fact, it has been shown \cite{Gonzalez-Jimenez:2021ohu} that using this method for $^{16}$O in the relativistic plane-wave impulse approximation (RPWIA) yields results within few percent of the fully factorized spectral function calculation~\cite{BENHAR1994493, PhysRevD.72.053005, RevModPhys.80.189}.

\subsection{\label{subsec:2-2}Final state interaction (FSI) models}
	In the following we will describe several options to include FSI in our results, {\it i.e.} different methods to calculate $\overbar{\Psi}^{s_N}\left(\mathbf{p_N},\mathbf{r}\right)$ on \eq{eq:current}, all of them within a fully relativistic and quantum framework.

	\begin{description}
		\item[Energy-dependent RMF (ED-RMF)] For this model, the nucleon ejected in the final state is represented by a scattering solution of the Dirac equation with the same RMF potential used to describe the initial nucleus but multiplied by a phenomenological function that weakens the potential for increasing nucleon momenta~\cite{PhysRevC.101.015503,PhysRevC.100.045501}. This model preserves orthogonality by construction because for the kinematics for which the overlap between the initial- and final-nucleon wave functions is significant, the initial and final mean-field potentials are the same.
						
		\item[Relativistic Optical Potential (ROP)] The ejected nucleon moves across the residual hadronic system under the influence of a phenomenological relativistic optical potential fitted to reproduce elastic proton-nucleus scattering data~\cite{PhysRevC.47.297, PhysRevC.80.034605} in the context of the optical model. This potential contains a real and an imaginary term, where the latter accounts for loses to inelastic channels. Thus, the ROP describes scenarios where the ejected nucleon propagates through the residual nucleus suffering only elastic scattering and consequently no other hadrons are created in the process. Hence, the ROP describes a contribution to the situation where only one proton and no other hadrons are detected in the final state, although additional hadrons can appear due to MEC or to initial state correlations, if the missing energy is large enough. The RDWIA-ROP approach has been used in the past to describe exclusive electron scattering $\left(e,e'p\right)$~\cite{PhysRevC.48.2731, PhysRevC.51.3246, PhysRevC.64.024614} experiments, for which a missing energy below the two-nucleon knock-out threshold can be determined from the detection of the final electron and proton in coincidence, plus the knowledge of the energy of the initial electron. In case of neutrino scattering, however, due to the fact that the energy distribution of the neutrino beams is very wide, the mere detection in coincidence of a muon and a proton in the final state does not guarantee control of the missing energy. Thus, the measured events would be composed of contributions beyond the elastic one described by the ROP. A simple way to consider in the final state the events beyond the elastic channel is to take only the real part of the ROP ($\mathbf{rROP}$), that is, removing the absorption into the elastic-only channel. This has been shown to be quite effective in describing inclusive cross-section measurements~\cite{PhysRevC.101.015503,PhysRevC.100.045501} which include all hadronic final states, both elastic and inelastic channels. Both rROP and  ED-RMF models  do not include losses to inelastic channels ({\it i.e.} both are real potentials) and are consistent with special relativity and quantum mechanics, although the orthogonality between the initial and final state is not as good for the rROP model as for the ED-RMF model~\cite{PhysRevC.100.045501}. Consequently, differences between them are expected to be found for relatively small momentum of the proton where orthogonality becomes an issue. The ROP potential used in this work is the so-called energy-dependent A-independent carbon (EDAI-C) potential~\cite{PhysRevC.47.297}.
	
		\item[Relativistic Plane-Wave IA (RPWIA)] For this model the ejected nucleon is described by a relativistic plane wave. Therefore, in this case, FSI are neglected. We include this model in our study to assess the importance of FSI in the description of semi-inclusive processes. 
	\end{description}

\section{\label{sec:3}SuSAv2 implementation in GENIE}

	Currently, many models in neutrino event generators are only able to calculate inclusive cross sections, {\it i.e.} cross sections that are function of the final lepton kinematics, where an integration over the hadronic final states is assumed. Therefore, these models can be used to predict the kinematics of the outgoing leptons, but cannot be directly applied to the description of hadrons in the final state. Nevertheless, it is possible to generate semi-inclusive predictions, {\it i.e.} a lepton and a hadron in the final state, using inclusive models implemented in generators by using approximations such as the factorization approach~\cite{PhysRevD.101.033003}. Since the description of these semi-inclusive reactions has an impact on the oscillation analyses, it is imperative to test the validity of these approximations against experimental measurements of cross sections and also against microscopic neutrino interaction models that can predict final lepton and hadron kinematics without approximations. 
	
	Among the different nuclear models for neutrino interactions, those based on the RMF theory are promising candidates to be implemented in event generators due to their accurate description of the nuclear dynamics and their good agreement with both inclusive and semi-inclusive electron- and neutrino-nucleus scattering data without relying on any factorization approach. As a first attempt in this direction, the SuSAv2-MEC model \cite{PhysRevC.90.035501, PhysRevD.94.013012, PhysRevC.99.042501}, a purely inclusive approach based on the RMF theory which has proven to successfully predict inclusive cross sections for electrons and neutrinos in a wide range of kinematics, has been recently included in the neutrino event generator GENIE \cite{PhysRevD.101.033003} for both 1p1h and 2p2h channels. This constitutes a first step for the implementation of the more sophisticated RMF models in further works and also allows to test factorization approaches.
	
	This implementation has been carried out via SuSAv2 1p1h and 2p2h hadron tensor tables, $H_{\mu\nu}(q,\omega)$, using a binning of 5 MeV in the transferred momentum and energy which is combined with GENIE's interpolation methods between adjacent bins. A factorization approach is assumed to generate the outgoing hadronic state where the initial state nucleon momentum is chosen by independently sampling from a local (global) Fermi gas nuclear model for the 1p1h (2p2h) channel. 
	
	For the 1p1h channel, the transferred energy $\omega$ is then reduced to take into account the removal energy of the nucleon based on a momentum-transfer dependent SuSAv2 analysis. The momentum transferred to the nucleon is altered so that the outgoing nucleon is on-shell, assuring momentum conservation by giving the appropriate amount to the residual nucleus. Within the impulse approximation approach, the energy transfer predicted from the inclusive interaction is initially given entirely to a single nucleon in the target and none to the residual nucleus. Finally, the propagation is carried out through the nucleus using GENIE's cascade FSI model.
	
	In the case of the 2p2h channel, based on the relativistic Fermi gas (RFG) calculation ~\cite{RuizSimo:2016rtu,RuizSimo:2016rqz}, a constant energy value is removed from the cluster of two nucleons to consider the removal energy. The probability of having neutron-neutron (proton-proton for antineutrinos) or proton-neutron pairs as initial cluster is chosen based on the kinematics of the inclusive interaction using the SuSAv2-MEC 2p2h theoretical model~\cite{RuizSimo:2016ikw}. The transferred momentum and energy are shared equally between the cluster components, one neutron (proton) is turned into a proton (neutron) for neutrinos (antineutrinos) and the cluster breaks up into two nucleons. The two nucleons are then propagated through the nucleus via GENIE's ``hN'' cascade FSI model~\cite{Dytman:2021ohr}. 
	
	The implementation for both 1p1h and 2p2h contributions has been widely validated against the original models for inclusive measurements.

\section{\label{sec:piabs}Pion absorption from GENIE}
	GENIE can further be used to model the pion absorption contribution to CC0$\pi$ measurements. The predominant contribution stems from GENIE's simulation of single pion production using the Berger-Sehgal model~\cite{Berger:2007rq}, which produces pions and nucleons which are propagated through the nucleus via the same ``hN'' FSI used for other interaction channels. In some fraction of the events the outgoing pions are absorbed within the nuclear medium (typically ejecting additional nucleons in the process). There is additionally some small contribution from more inelastic channels whose mesons are all absorbed by FSI.
		
\section{\label{sec:4}Results}

	We now proceed to compare all the available semi-inclusive CC0$\pi$Np cross-section measurements for T2K and MINER$\nu$A with the predictions of two approaches for the 1p1h sector: the RMF model based on \eq{semi-inclusive}, where FSI are implemented using the different prescriptions described in Sec.~\ref{subsec:2-2}, and the 1p1h GENIE-SuSAv2 implementation described in Sec.~\ref{sec:3}. For both approaches we add on top the SuSAv2-2p2h MEC and pion absorption contributions calculated with GENIE. The processing of GENIE output and its comparison to experimental data was made using the NUISANCE framework~\cite{Stowell_2017}.
	
	For MINER$\nu$A~\cite{PhysRevLett.121.022504, PhysRevD.101.092001} the comparison is made as function of the muon and proton kinematics and as function of the transverse kinematic imbalances (TKI)~\cite{PhysRevC.94.015503, dolan2018exploring} that measure correlations between the final muon and proton in the plane transverse to the neutrino direction. Additionally, for T2K we show the cross sections as function of the so called inferred variables (IV) that compare the momentum and angle of the ejected proton with the proton kinematics inferred from the measured final state muon kinematics when assuming a QE interaction on a target nucleon at rest~\cite{PhysRevD.98.032003}. The specific experimental constraints applied to T2K and MINER$\nu$A measurements are summarized in Table~\ref{table:T2K constrains} and Table~\ref{table:Minerva constrains}, respectively.

	\begin{table*}[!htbp]%[!b]
	\centering
	\begin{tabular}{cccccccccccccccc}
		\hline
		\toprule\toprule
		\Mark{T2K}              &  &   $k'$  &   & & $\cos{\theta_l}$ & &    &  $p_N$  &  && $\cos{\theta_N^L}$& && $\phi_N^L$ &\\\midrule
		TKI               & & $> 0.25$ GeV & &  &      $> -0.6$   &  &  &0.45-1.0 GeV& &  &    $> 0.4$    &&&-&  \\ \midrule
		IV                &  &    -   &    &   &       -         & &  &$> 0.45$ GeV& & &     $> 0.4$      &&&-&\\ 
		\bottomrule\bottomrule
		\hline
	\end{tabular}
	\caption{\label{table:T2K constrains}Phase-space restrictions applied to the CC0$\pi$ cross-section measurements with one muon and at least one proton in the final state shown by T2K collaboration in \cite{PhysRevD.98.032003}. }
	\end{table*}

	\begin{table*}[!htbp]%[!b]
		\centering
			\begin{tabular}{cccccccccccccccc} 
				\hline
				\toprule\toprule
				\Mark{MINER$\nu$A}              &  &   $k'$  &   & & $\cos{\theta_l}$ & &    &  $p_N$  &  && $\cos{\theta_N^L}$& & &$\phi_N^L$&\\\midrule
				All analyses               & & 1.5-10 GeV & &  &      $> 0.939$   &  &  &0.45-1.2 GeV& &  &    $> 0.342$    & &&-& \\ 
				\bottomrule\bottomrule
				\hline
			\end{tabular}
		\caption{\label{table:Minerva constrains}Phase-space restrictions applied to the CC0$\pi$ cross-section measurements with one muon and at least one proton in the final state shown by MINER$\nu$A collaboration in \cite{PhysRevD.101.092001,PhysRevLett.121.022504}.}
	\end{table*}

\subsection{\label{subsec:4-1}T2K}

	\subsubsection*{CC0$\pi$0p}
	\begin{figure*}[!hptb]
		\centering
		\makebox[\textwidth]{\includegraphics[width=\textwidth,height=0.70\paperheight]{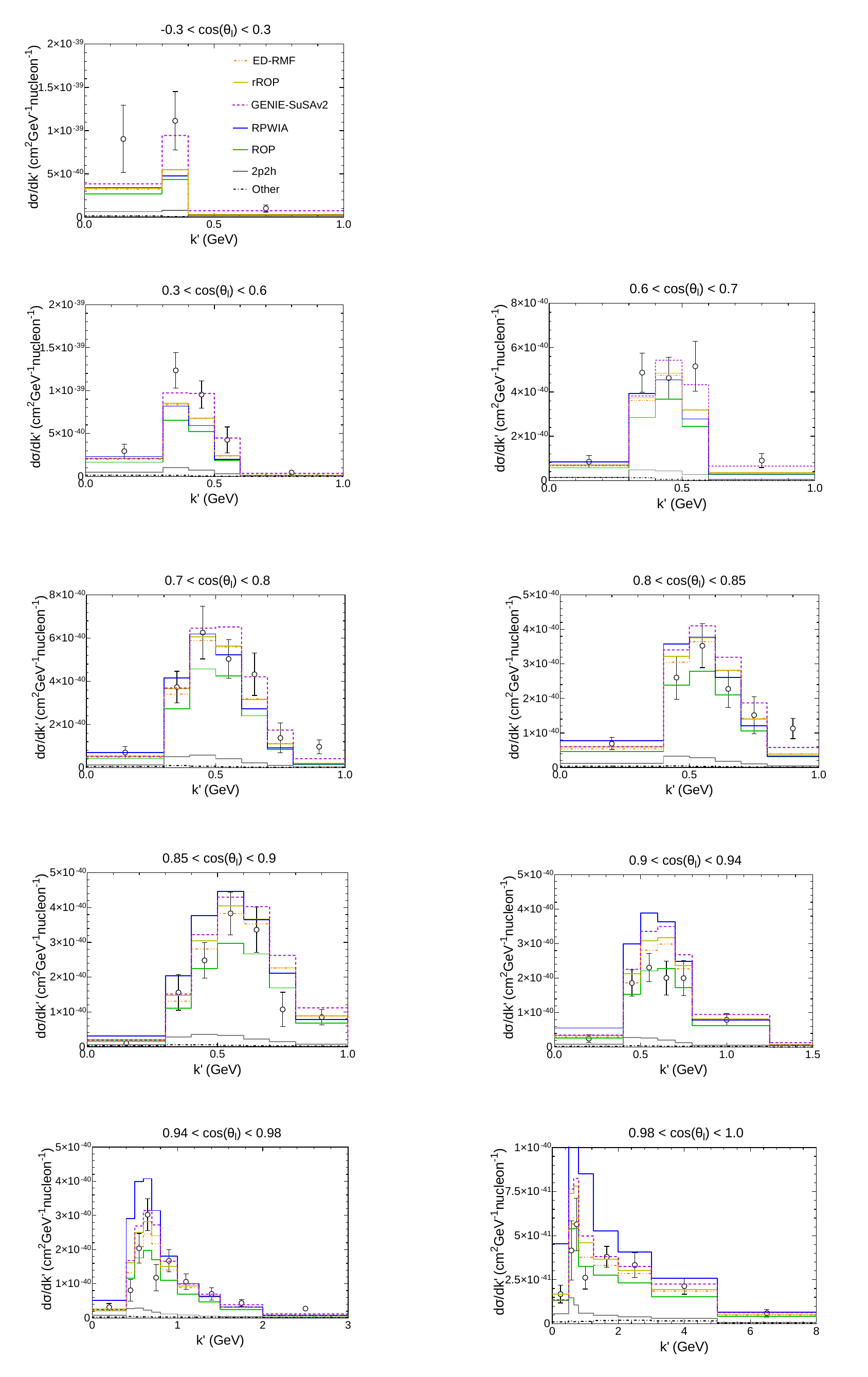}}
		\caption{\label{fig:T2K inclusive} T2K CC0$\pi$ semi-inclusive $\nu_\mu-^{12}$C cross section without protons in the final state with momenta above 0.5 GeV as function of final muon kinematics. All curves include the 2p2h and pion absorption contributions (also shown separately), evaluated using GENIE. Cross-section measurements taken from \cite{PhysRevD.98.032003}.}
	\end{figure*}
	In Fig.~\ref{fig:T2K inclusive} we compare the microscopic calculations and the GENIE implementation of SuSAv2 with T2K CC0$\pi$ cross-section measurements without protons in the final state with momenta above 0.5 GeV (CC0$\pi$0p) as function of the final muon kinematics. For backward angles the microscopic calculation predicts a rather small difference between the results for RPWIA and the models with FSI (rROP, ROP and ED-RMF), all of them underestimating the experimental measurements in contrast with the better agreement achieved with GENIE-SuSAv2. As we move to more forward angles GENIE-SuSAv2 predictions start to overestimate some of the experimental points, an outcome probably due to scaling/factorization violations and poor treatment of low-energy effects which are accounted for more consistently in ED-RMF. As discussed before, at very forward angles orthogonalization issues are important, yielding spurious contributions to the cross section for the RPWIA model, which largely overestimate the data at low values of $k'$. For all the angular bins the 2p2h, although non-negligible, is limited to a few percent of the total cross section. Interestingly, the final state proton kinematic restriction ($p_N <$ 0.5 GeV) leaves the pion absorption contribution negligible. No model is able to reproduce the sharp oscillation shown by the data just after the maximum in the last two bins (0.94 $< \cos{\theta_\mu} <$ 0.98 and 0.98 $< \cos{\theta_\mu} <$ 1.0), but it should be noted that, once the reported correlations in the measured cross section are accounted for, the measurement shows no significant preference for an oscillation in the cross section. To quantify the agreement of the different models with the measurements, in Appendix~\ref{appendix} we include a $\chi^2$ analysis using the covariance matrices provided with the cross sections measurements. The results for T2K are summarized in Table~\ref{table: chi2 t2k}. For CC0$\pi$ measurements without protons in the final state with momenta above 0.5 GeV the ROP and ED-RMF models have associated a smaller $\chi^2$ compared with GENIE-SuSAv2 results, although still much larger than the number of degrees of freedom (d.o.f), indicating that low momentum protons are not quantitatively described by these models.

	\subsubsection*{CC0$\pi$Np}
\begin{figure*}[!htbp]
		\centering
		\includegraphics[width=\textwidth,height=0.70\paperheight]{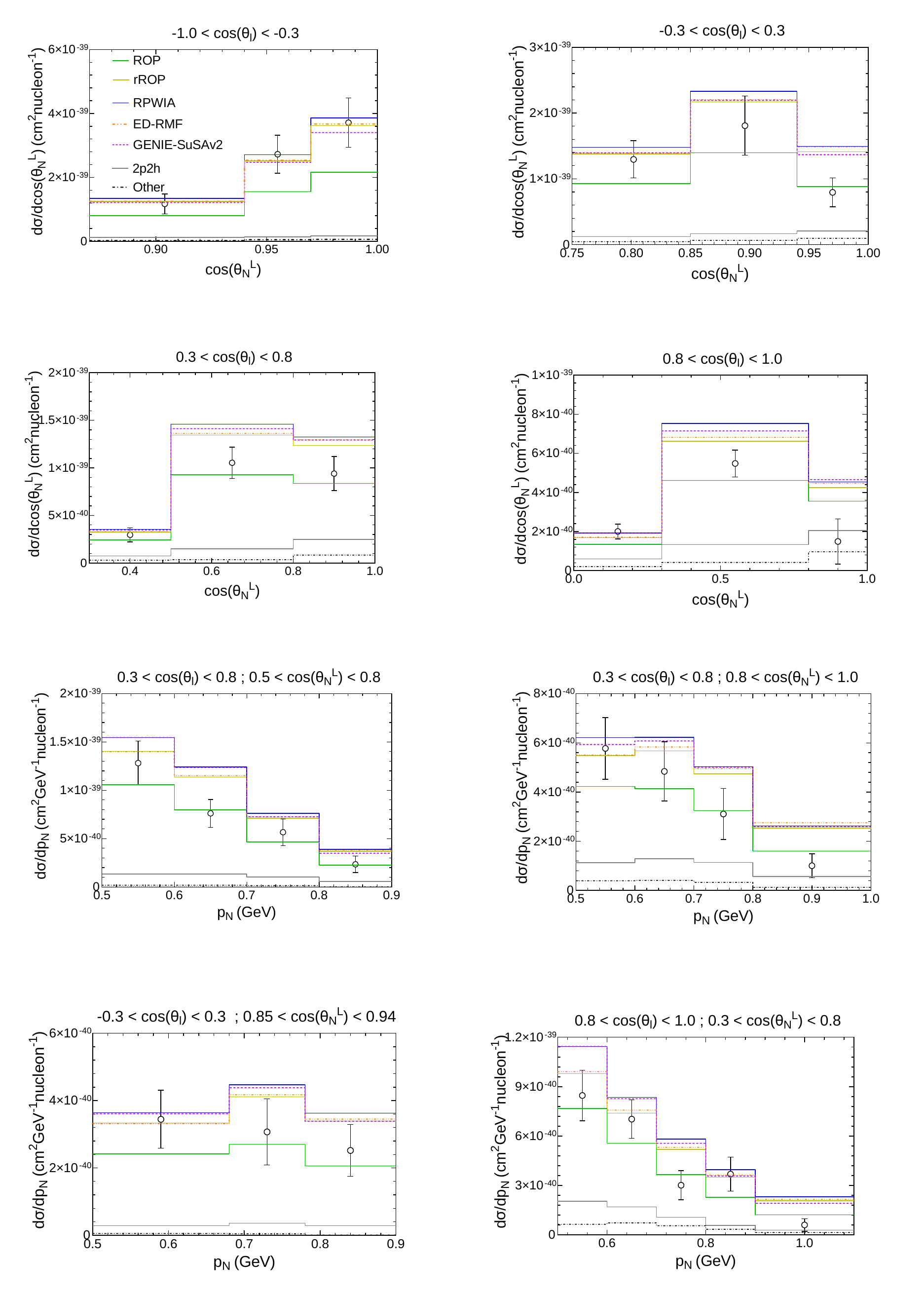}
		\caption{\label{fig:T2K semi-inclusive} T2K CC0$\pi$ semi-inclusive $\nu_\mu-^{12}$C cross section with protons in the final state with momenta above 0.5 GeV as function of the final proton and muon kinematics. All curves include the 2p2h and pion absorption contributions (also shown separately), evaluated using GENIE. Cross-section measurements taken from \cite{PhysRevD.98.032003}.}
	\end{figure*}

\begin{figure}[!htbp]
		\centering
		\includegraphics[width=0.5\textwidth]{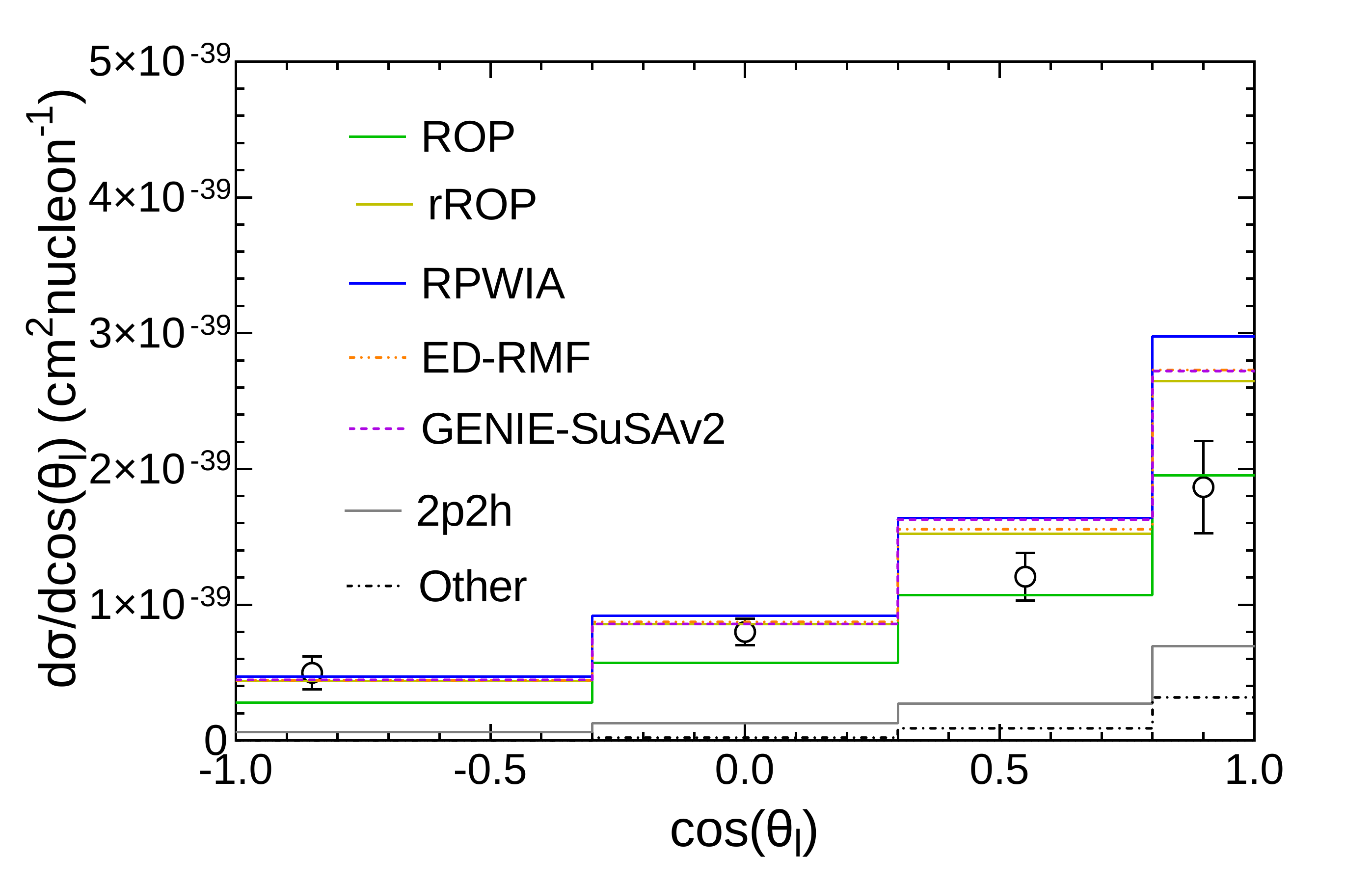}
		\caption{\label{fig:T2K_angle_semi-inclusive} T2K CC0$\pi$ semi-inclusive $\nu_\mu-^{12}$C cross section with protons in the final state with momenta above 0.5 GeV as function of the muon scattering angle. All curves include the 2p2h and pion absorption contributions (also shown separately), evaluated using GENIE. Cross-section measurements taken from \cite{PhysRevD.98.032003}.}
	\end{figure}

	In Fig.~\ref{fig:T2K semi-inclusive} and Fig.~\ref{fig:T2K_angle_semi-inclusive} the different models are compared with T2K semi-inclusive CC0$\pi$Np cross-section measurements with protons with momenta above 0.5 GeV as function of the leading proton's kinematics and the muon scattering angle, respectively. In general, the 2p2h channel seems to be more relevant for this case, especially for forward scattering angles. This is not surprising, since the 2p2h cross section is peaked at higher $\omega$ (hence higher $p_N$) than the quasi-elastic cross section. The pion absorption channel also appears more relevant, but only in specific regions of outgoing lepton and nucleon kinematics (at relatively forward lepton \textit{and} nucleon scattering angles). As a consequence the data with $p_N >$ 0.5 GeV are more affected by non-quasielastic contributions than at $p_N <$ 0.5 GeV. The GENIE-SuSAv2 results slightly overestimate some of the experimental points, while the ED-RMF and rROP models tend to match or improve the agreement, especially for proton momentum around 0.5-0.7 GeV. It is interesting to note that the ROP model in Fig.~\ref{fig:T2K_angle_semi-inclusive} describes better the cross section measurements for $ 0.3 <\cos{\theta_l} < 1.0$ than the rest of the models, but the situation reverses for $-1.0 <\cos{\theta_l} < 0.3$, with ROP underestimating the cross section measurements. ROP predicts the cross section corresponding to the case in which the struck nucleon interacts only elastically with the residual nucleus, i.e., it does not knock out other nucleons or create new mesons in its way out. Thus, if one would not include the background contribution due to short-range correlations, that appears at large $E_m$-$p_m$ (see Fig.~\ref{fig:rho}), and that necessarily corresponds to a process with at least two nucleons in the final state, then the ROP model gives a lower bound estimate of the one and only one proton, and no other hadron, in the final state. We would expect in general that the experimental measurements are above the ROP predictions, consistently with the fact that the experimental signal includes more channels than the one represented in the ROP, namely, that the nucleon knocked out by the neutrino just interacts elastically while traveling off the nucleus

	The $\chi^2$ comparison presented in Table~\ref{table: chi2 t2k} shows good agreement of the ROP results with the measurements with a $\chi^2/d.o.f$ close to 1. As shown in Appendix~\ref{appendix}, this strong preference for the ROP model is driven up by bins with high proton momentum in the cross sections as function of $p_N$ with 0.3 $< \cos{\theta_N^L} <$ 0.8 and 0.8 $< \cos{\theta_N^L} <$ 1.0.

	\subsubsection*{Inferred variables}
	
\begin{figure*}[!htbp]
		\centering
		\includegraphics[width=\textwidth,height=0.70\paperheight]{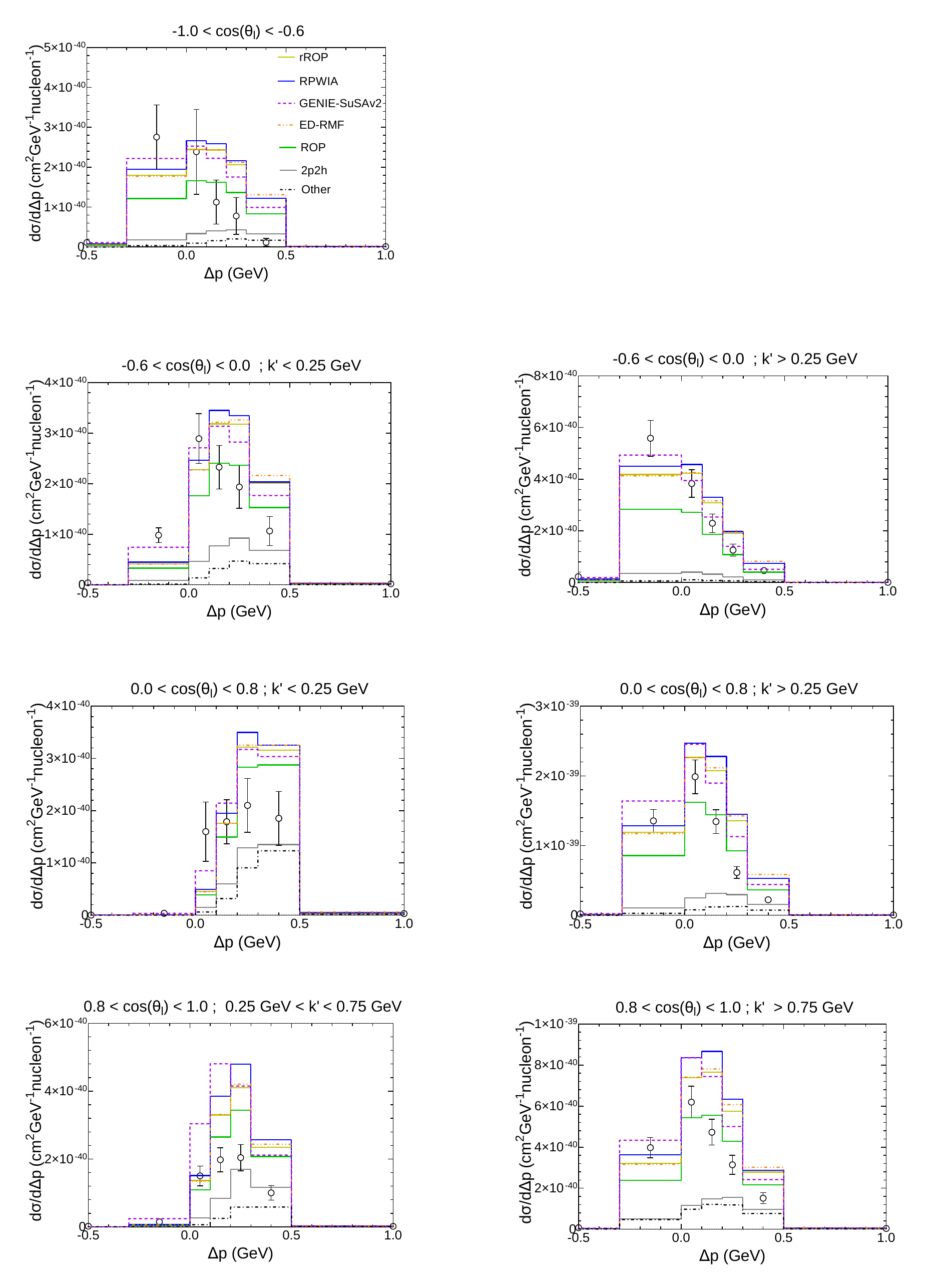}
		\caption{\label{fig:IV deltap} T2K CC0$\pi$ semi-inclusive $\nu_\mu-^{12}$C cross section as function of the variable $\Delta p$ defined in \eq{inferred variables} for different muon kinematic bins with constrains of the proton kinematics given in Table~\ref{table:T2K constrains}. All curves include the 2p2h and pion absorption contributions (also shown separately), evaluated using GENIE. Cross-section measurements taken from \cite{PhysRevD.98.032003}.}
	\end{figure*}

\begin{figure*}[!htbp]
		\centering
		\includegraphics[width=\textwidth,height=0.65\paperheight]{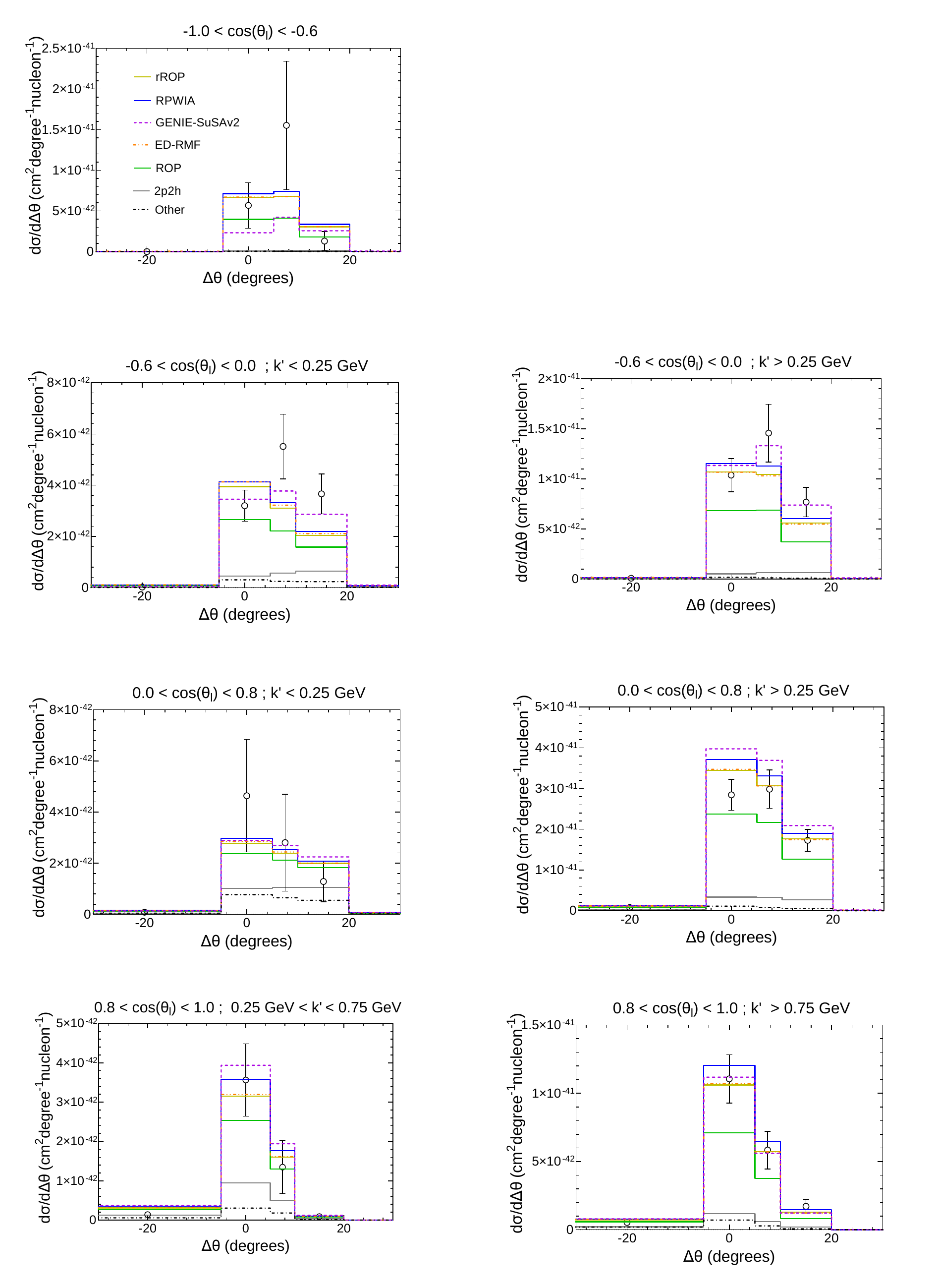}
		\caption{\label{fig:IV deltatheta} T2K CC0$\pi$ semi-inclusive $\nu_\mu-^{12}$C cross section as function of the variable $\Delta\theta$ defined in \eq{inferred variables2} in different muon kinematic bins with constrains of the proton kinematics given in Table~\ref{table:T2K constrains}. All curves include the 2p2h and pion absorption contributions (also shown separately), evaluated using GENIE. Cross-section measurements taken from \cite{PhysRevD.98.032003}. For readability, the axis range has been reduced to $\left[-30^\circ,+30^\circ\right]$ hiding an experimental bin above $30^\circ$ with very low cross section and centering the $\left[-360^\circ,-5^\circ\right]$ experimental bin around $-20^\circ$.
}
	\end{figure*}

\begin{figure*}[!htbp]
		\centering
		\includegraphics[width=\textwidth,height=0.70\paperheight]{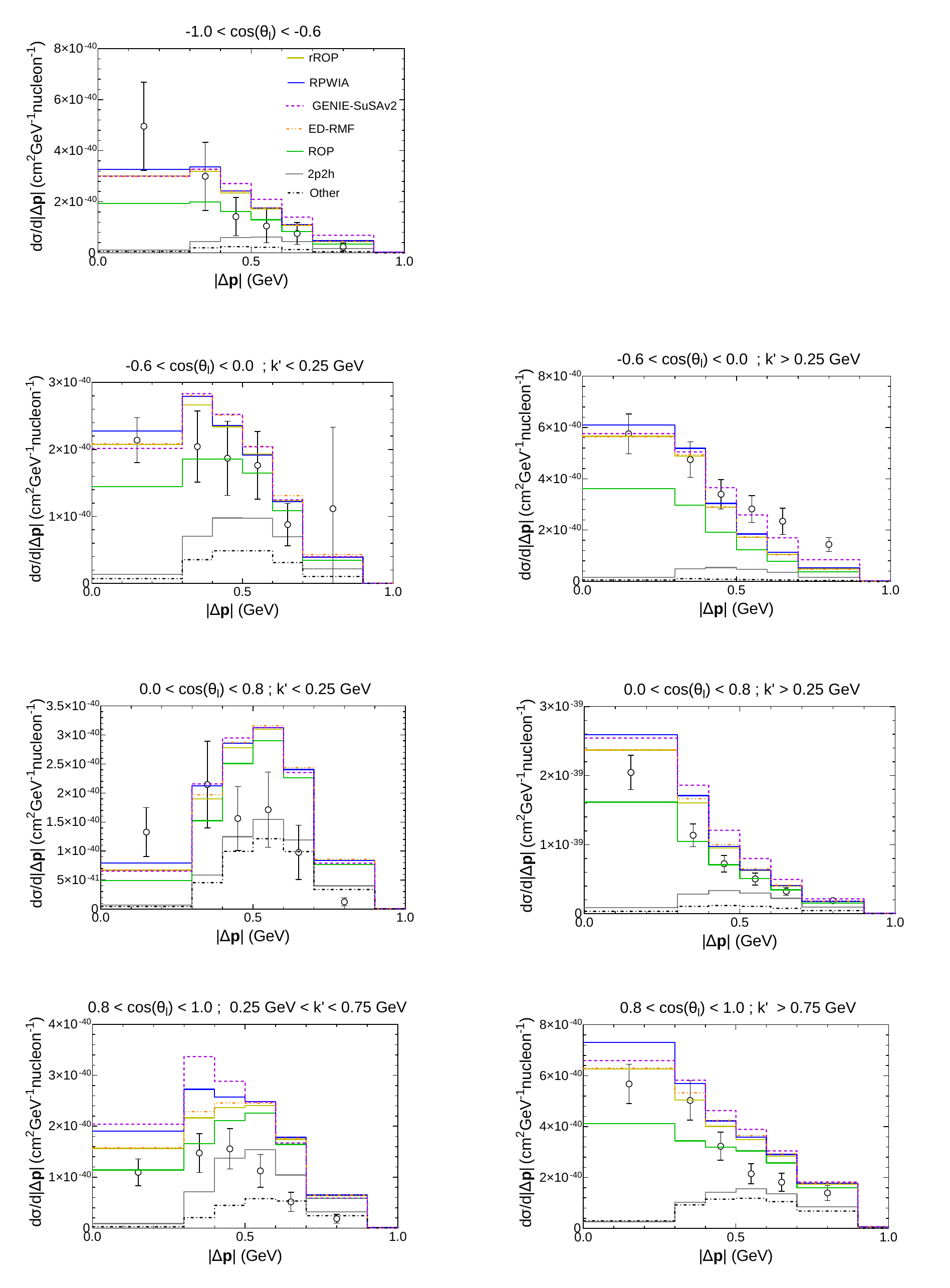}
		\caption{\label{fig:IV deltamodp} T2K CC0$\pi$ semi-inclusive $\nu_\mu-^{12}$C cross section as function of the variable $\left|\Delta\mathbf{p}\right|$ defined in \eq{inferred variables3} in different muon kinematic bins with constrains of the proton kinematics given in Table~\ref{table:T2K constrains}. All curves include the 2p2h and pion absorption contributions (also shown separately), evaluated using GENIE. Cross-section measurements taken from \cite{PhysRevD.98.032003}.}
	\end{figure*}

	In \Cref{fig:IV deltap,fig:IV deltatheta,fig:IV deltamodp} we show the results as function of the proton inferred kinematics variable (IV) for the different models. The IV variables are defined as~\cite{PhysRevD.98.032003}
	\begin{eqnarray}\label{inferred variables}
		\Delta p&=& \left| {\bf p_N} \right| -  \left| {\bf p_N}^{\rm inf} \right|
			\\ \label{inferred variables2}
		\Delta\theta&=& \arccos \left({\bf \hat p_N}^{\rm inf}\cdot {\bf \hat z}\right)	\\\label{inferred variables3}
		\left|\Delta \mathbf{p}\right| &=& \left|{\bf p_N} - {\bf p_N}^{\rm inf} \right|\,,
	\end{eqnarray}
	where ${\bf \hat z}$ denotes the neutrino beam direction and ${\bf p_N}^{\rm inf} = {\bf k_\nu}^{\rm inf}-{\bf k'}$ is the final proton momentum inferred under the hypothesis that the neutrino interacts with a neutron at rest having mass $\tilde m_n = m_n-E_b$ (with $E_b$=25 MeV for carbon), namely
	\begin{equation}
		{\bf k_\nu}^{\rm inf} = \frac{m_p^2-m_\mu^2+2 E_l \tilde m_n-\tilde m_n^2}{2 \left(\tilde m_n - E_l+k'\cos\theta_l\right)} \ {\bf \hat z}	\,.		
	\end{equation}
			
	Based on the results of the GENIE-SuSAv2 2p2h model and GENIE's pion absorption predictions there are angular bins with areas heavily dominated by non-quasielastic channels, especially for the cross sections as function of $\Delta p$ and $\left|\Delta \mathbf{p}\right|$ in bins with small scattering angle and low momentum of the muon. For the $\left|\Delta \mathbf{p}\right|$ distribution there is a clear preference to require significant non-quasielastic contributions in the high momentum imbalance tail in the higher lepton momentum, intermediate lepton scattering angle slices, where the microscopic calculation shows small FSI effects by comparing the RPWIA results with the ED-RMF and rROP predictions. Regarding the comparison of the different 1p1h predictions, the biggest differences between the GENIE-SuSAv2, the ED-RMF and rROP microscopic results can be found for forward angles and low muon momentum, especially in the $\Delta p$ and $\left|\Delta \mathbf{p}\right|$ cross sections, where the GENIE-SuSAv2 estimation can be up to 50$\%$ higher than the ED-RMF result. This might be caused by the limitations of SuSAv2 model to describe correctly low-energy nuclear effects and scaling violations in the forward region. Even with this severe reduction compared with the results from GENIE-SuSAv2, the ED-RMF and rROP models still overestimate the cross-section measurements in these forward angles and/or low momentum bins due to a large contribution coming from non-quasielastic channels. This might be related to an overestimation of the 2p2h contribution associated to the extrapolation performed in GENIE to connect the inclusive 2p2h hadronic tensor evaluated microscopically to the semi-inclusive one used to simulate these cross sections. The disagreement may eventually be resolved by performing a fully semi-inclusive calculation where both the leptonic and hadronic variables are consistently handled. Notice that the agreement with the cross-section measurements is improved in the bin with forward angles and high muon momentum ($k'>0.75$ GeV) at low $\left|\Delta \mathbf{p}\right|$, but that the non-quasielastic contribution at higher $\left|\Delta \mathbf{p}\right|$ seems to remain too large. It is also interesting to note that, for the $\Delta\theta$ cross section in the bin with the most backward-going muons, the GENIE-SuSAv2 prediction falls below cross-section measurements and is even lower than the ROP estimation around zero imbalance, which might indicate too strong FSI. The $\chi^2$-values shown in Table~\ref{table: chi2 t2k} are large compared with the d.o.f for the three inferred variables, with the worse agreement obtained for the $\Delta\theta$ distribution. For this specific case, as explained in Appendix~\ref{appendix}, large contributions to $\chi^2$ come from two specific bins with very small cross section. If those bins are removed, agreement of the models based on RDWIA with the measurements is matched (rROP and ED-RMF) or improved (ROP) with respect to the GENIE-SuSAv2 model.
	
	\subsubsection*{TKI}
\begin{figure}[!htbp]
		\captionsetup[subfigure]{labelformat=empty} 
		\centering
		\subfloat[]{{\includegraphics[width=0.49\textwidth]{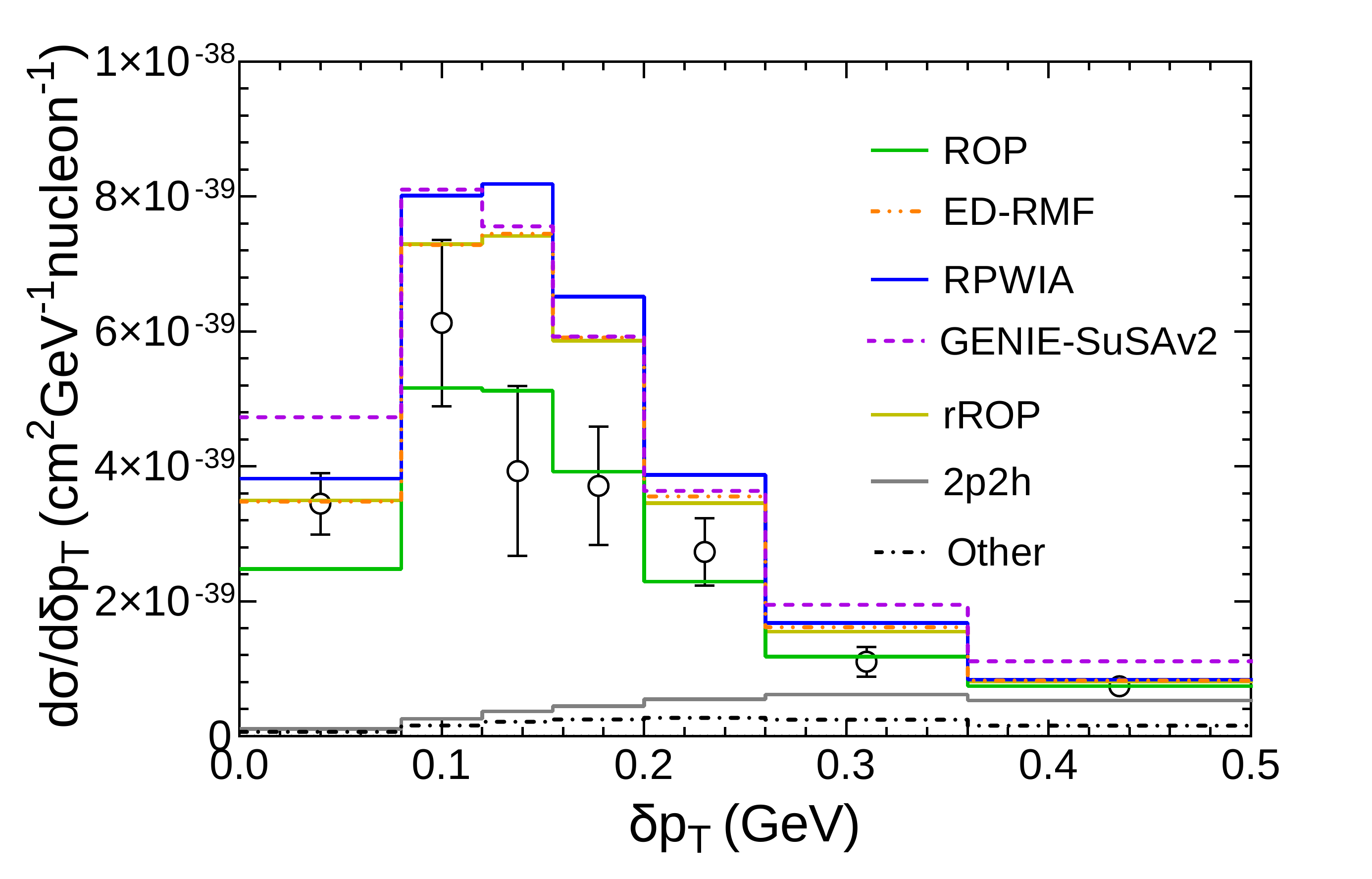}}}%
		\quad
		\subfloat[]{{\includegraphics[width=0.49\textwidth]{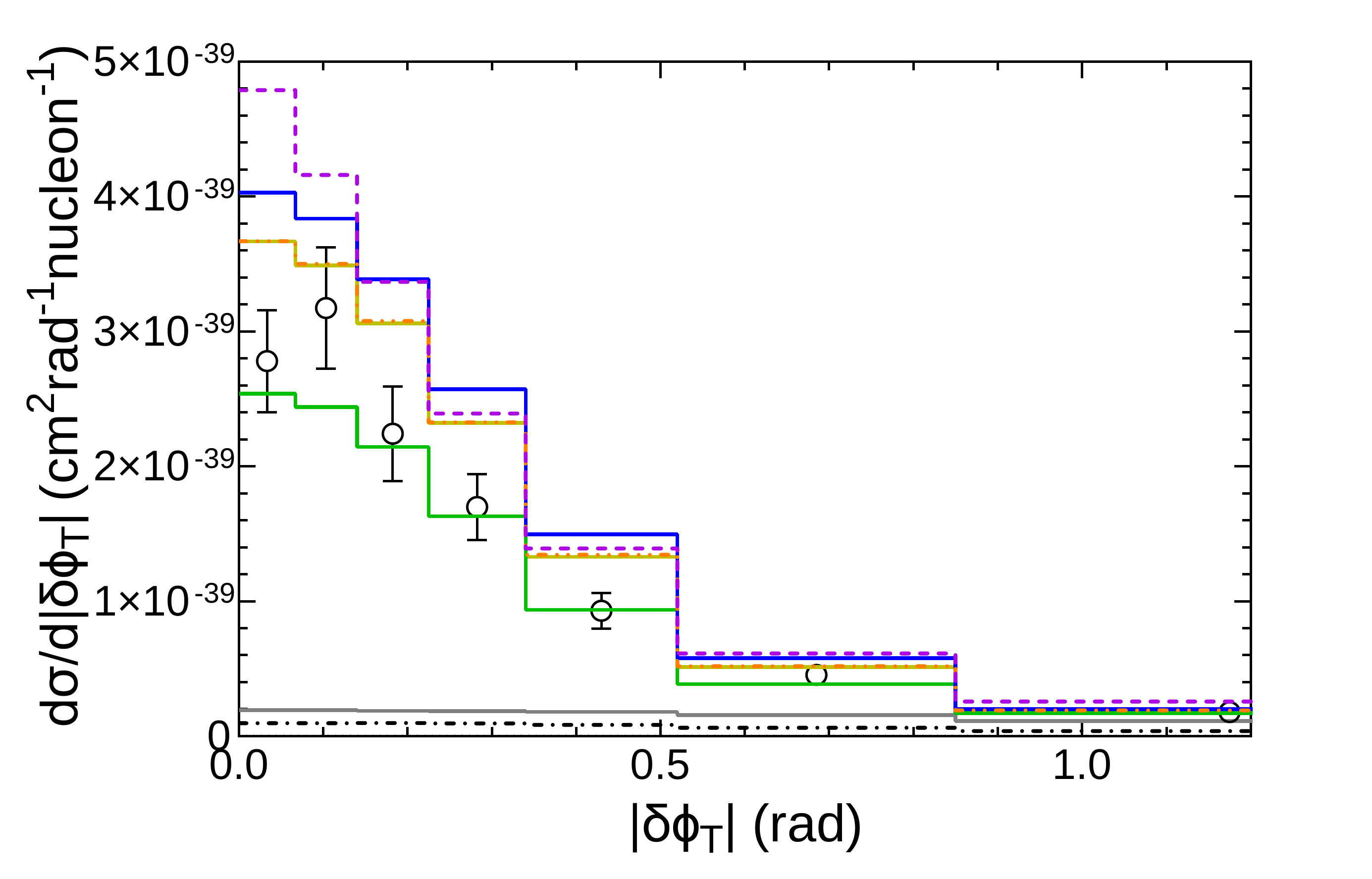}} }%
		\quad
		\subfloat[]{{\includegraphics[width=0.49\textwidth]{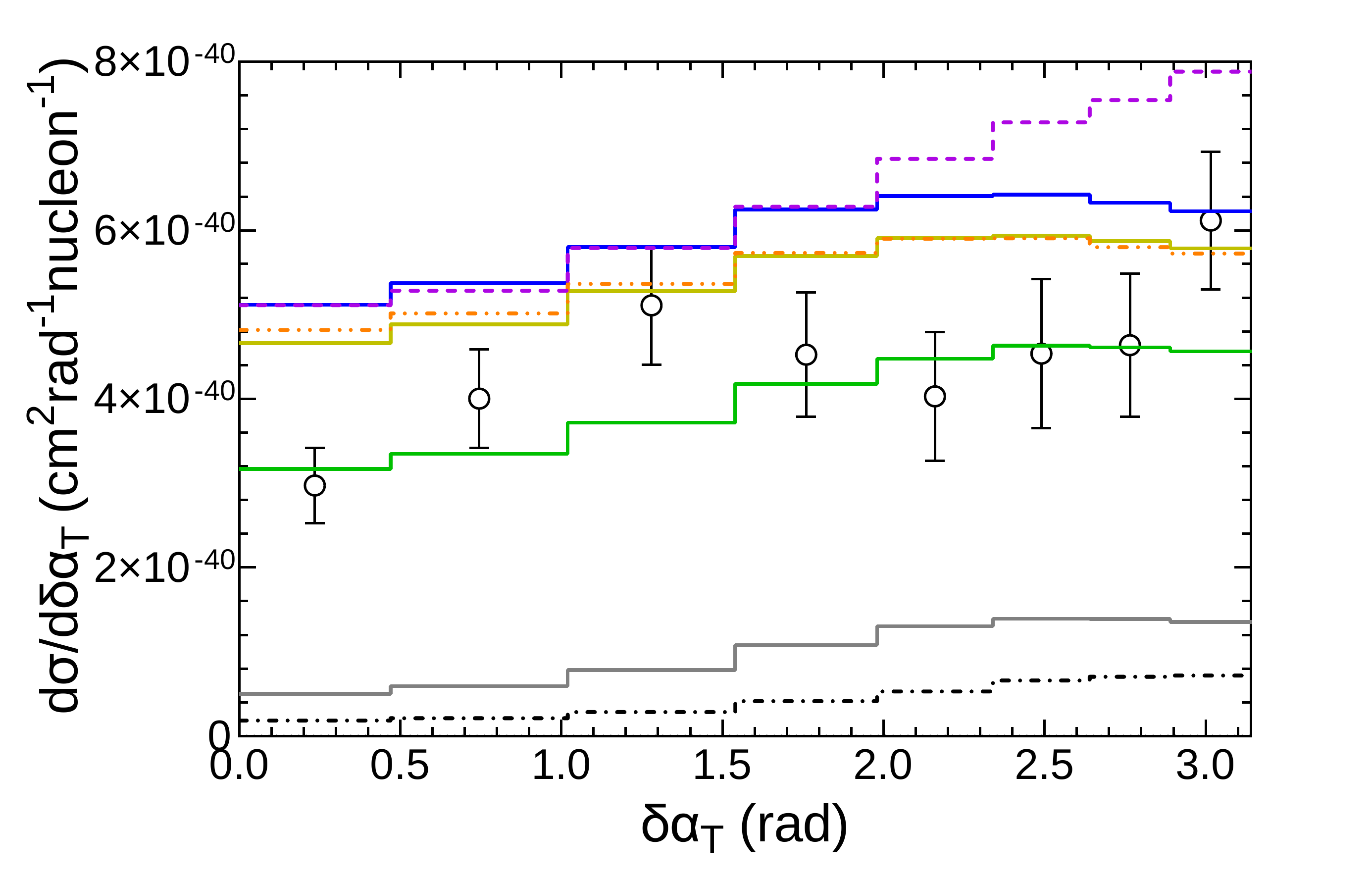}}}%
		\caption{\label{fig:T2K semi-inclusive STV} T2K CC0$\pi$ semi-inclusive $\nu_\mu-^{12}$C cross sections as function of the transverse kinematic imbalances $\delta p_T$, $\delta\alpha_T$ and $\left|\delta\phi_T\right|$ defined in Eqs.~(\ref{tki:dpt}-\ref{tki:dphit}). All curves include the 2p2h and pion absorption contributions (also shown separately), evaluated using GENIE. Cross-section measurements taken from~\cite{PhysRevD.98.032003}.}	
	\end{figure}

	The comparison of the cross sections as function of the transverse kinematic imbalances  for the different models with T2K measurements is presented in Fig.~\ref{fig:T2K semi-inclusive STV}. We recall here the definition of the TKI variables~\cite{PhysRevD.98.032003}
	\begin{eqnarray}\label{tki:dpt}
		\delta p_{T} &=& \left| {\bf \delta p_T}\right| =  \left| {\bf k'_T}+{\bf p_{N,T}}\right|\,,	\\\label{tki:dalphat}
		\delta\alpha_{T} &=& \arccos \left(- \frac {{\bf k'_T}\cdot{\bf \delta p_T}} {\left|{\bf k'_T}\right| \, \left|{\bf \delta p_T}\right| }\right)\,,
		\\\label{tki:dphit}
		\delta\phi_{T} &=&\arccos \left(- \frac{{\bf k'_T}\cdot{\bf  p_{N,T}}}{\left| {\bf k'_T} \right| \, \left| \bf  p_{N,T}\right| }\right)
			\,,
	\end{eqnarray}
	where the label $T$ refers to  projection on the plane transverse to the neutrino beam. In the absence of FSI and supposing a pure QE event, the momentum imbalance is generated entirely by the description of the initial nuclear state dynamics \cite{PhysRevC.94.015503,dolan2018exploring}. In this approximation $\delta p_T$ is a direct measurement of the transverse component of the bound nucleon momentum distribution, therefore the RFG model, widely used in neutrino event generators, would be at a disadvantage compared to more realistic nuclear models like the independent-particle shell model or the spectral function model \cite{PhysRevD.98.032003,dolan2018exploring}. This was explicitly shown in Ref.~\cite{PhysRevD.104.073008}, where the RFG was found to give a much poorer description of the $\delta p_T$ distribution than the shell model. The $\delta p_T$ distribution shown in Fig.~\ref{fig:T2K semi-inclusive STV} favours the ED-RMF and rROP calculations over the GENIE-SuSAv2 predictions in the low $\delta p_T$ region, which is mainly dominated by initial-state effects with negligible contribution from the 2p2h and pion absorption channels. This could be caused by the inconsistencies of the implementation of the SuSAv2 model, which is based on the RMF theory, in GENIE, that generates the initial state nucleon using a local Fermi gas model. For imbalances above the Fermi level, nucleon-nucleon correlations become more important and the microscopic calculation predicts small FSI effects. In this region all the microscopic models except the ROP model overestimate the cross-section measurements after including the 2p2h and pion absorption contributions calculated with GENIE, although the comparison is also inconsistent because the 2p2h contribution is calculated with a Fermi gas while the microscopic calculations for the quasielastic process use the RMF model with corrections to include nucleon-nucleon correlations. In any case, it is clear that the QE contribution with nucleon-nucleon correlations included is not enough to describe the region of high-momentum imbalance and additional contributions are essential to describe the experimental results. 

	Regarding the angular TKI, $\delta\phi_T$ is more dependent on the neutrino energy and less sensitive to nuclear effects than $\delta p_T$~\cite{PhysRevC.94.015503}. The variable $\delta\alpha_T$ measures with good approximation the angle between the initial nucleon momentum and the transferred momentum \cite{PhysRevC.94.015503}. All the model predictions except ROP as function of $\delta\phi_T$ shown in Fig.~\ref{fig:T2K semi-inclusive STV} overestimate the cross-section measurements, although the overestimation is less severe in the case of the ED-RMF and rROP models for low values of $\delta\phi_T$. In the case of $\delta\alpha_T$, it is expected to have a rather flat distribution due to the isotropy of the momentum distribution of the bound nucleon which is broken mainly by non-quasielastic effects and FSI, although this deviation from flatness is partially washed out by the constraint on the outgoing proton kinematics in current experimentally accessible signal definitions. However, it is interesting to note that all of the microscopic models, including RPWIA, predict a significant CCQE-driven rise in $\delta\alpha_T$ when there is no or low proton momentum threshold (although this is not shown here), in contrast to what is often predicted by neutrino event generators. In the results presented in Fig.~\ref{fig:T2K semi-inclusive STV} there is an overestimation of the cross-section measurements by all the models except the ROP, which is less significant for the microscopic calculations using the ED-RMF and rROP models compared with GENIE-SuSAv2 results. The non-quasielastic contributions become more relevant for higher values of $\delta\alpha_T$ where the biggest differences between the microscopic and the GENIE-SuSAv2 calculations also appear, with the former in better agreement with the cross-section measurements. The differences between both kinds of calculations could be explained by the different treatment of FSI. Simulations performed with the neutrino event generator NuWro~\cite{Niewczas:2019fro}, which also uses a semi-classical cascade model for FSI with tuned parameters, are in better agreement with the microscopic calculation than with GENIE-SuSAv2. Quantitatively, $\chi^2$ values summarized in Table~\ref{table: chi2 t2k} correspond to values of $\chi^2/d.o.f$ close to 1 for the ROP model for the three TKI variables, with the best agreement found for the $\delta\alpha_T$ distribution.

\subsection{\label{subsec:4-2}MINER$\nu$A}
	
\begin{figure*}[t]
		\centering
		\includegraphics[width=\textwidth]{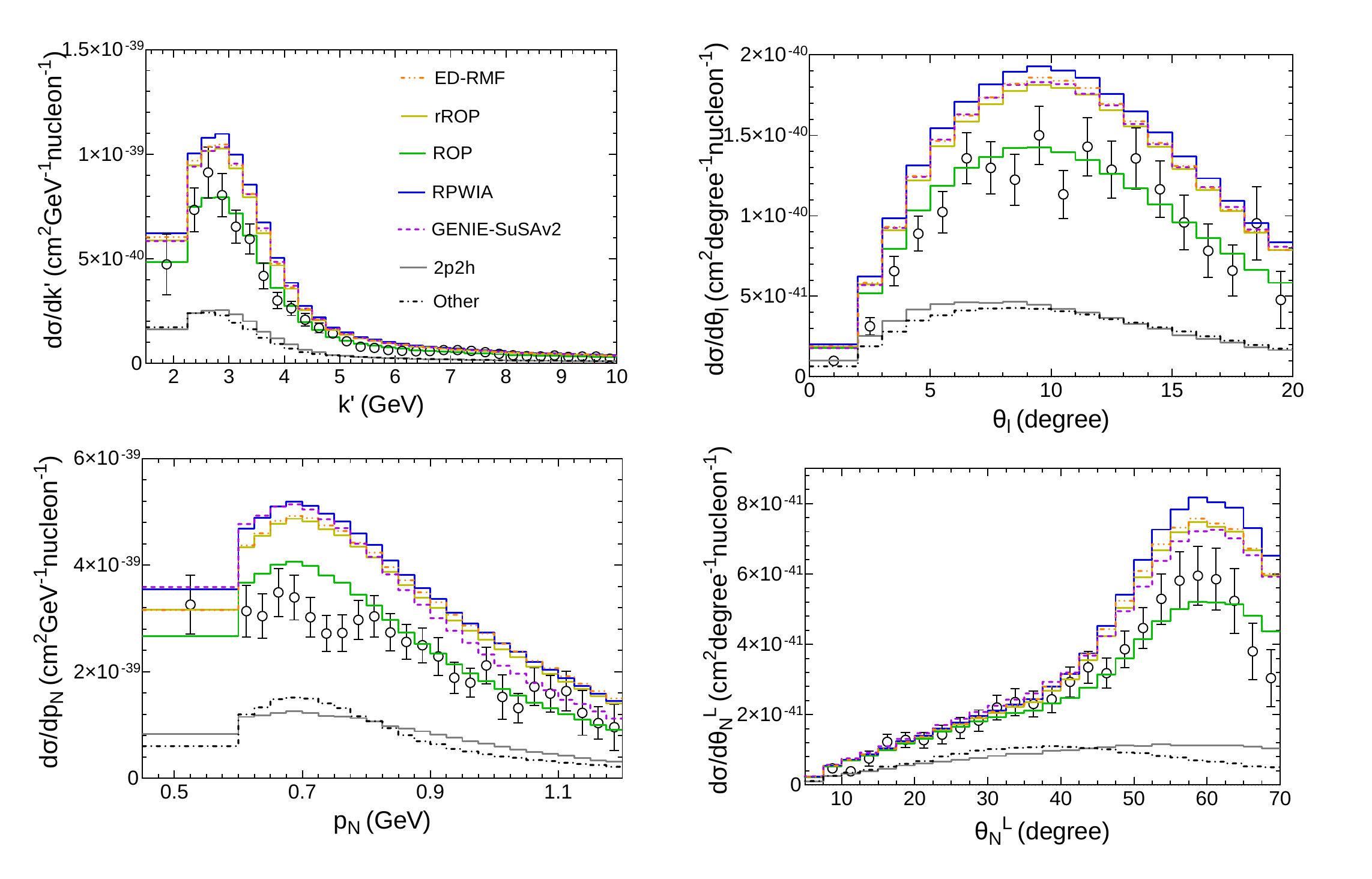}
		\caption{\label{fig:NV minerva} MINER$\nu$A semi-inclusive $\nu_\mu-^{12}$C cross section as function of the final muon momentum and scattering angle (top) and as function of the final proton momentum and polar angle (bottom). All curves include the 2p2h and pion absorption contributions (also shown separately), evaluated using GENIE. The original paper from MINER$\nu$A was \cite{PhysRevLett.121.022504} but the cross-section measurements shown here were taken from \cite{PhysRevD.101.092001} which corrected a mismodelling in GENIE's elastic FSI that affected the cross-section measurements presented in the first paper.}
	\end{figure*}

	We now compare our predictions with the results published by the MINER$\nu$A collaboration in Refs.~\cite{PhysRevLett.121.022504,PhysRevD.101.092001}. Despite a larger contribution from non-quasielastic channels, due to the higher energy neutrino beam ($\langle E_\nu\rangle$=3 GeV), the semi-inclusive cross sections predicted by ROP, and shown in Fig.~\ref{fig:NV minerva} as function of the muon and proton kinematics, seem qualitatively in reasonable agreement with experimental measurements, except for the $\theta_N^L$ cross section where there is an underestimation of the cross-section measurements for low values of $\theta_N^L$ and an overestimation in the high-$\theta_N^L$ region. This is partially confirmed by the $\chi^2$-values presented in Table~\ref{table: chi2 minerva}, which show ROP as the favored model which best matches the measurements of lepton kinematic variables. Whilst all models other than ROP overpredict the cross section, agreement between the ED-RMF, rROP and the GENIE-SuSAv2 predictions is very good except for the $p_N$ distribution where differences can be seen in the whole interval of proton momentum. It should be noted that the apparent overprediction of the non-ROP models may be due to a mismodelling of the strength of the 2p2h or pion absorption contributions and may therefore not suggest an issue in the CCQE modelling.

\begin{figure}[!htbp]
		\captionsetup[subfigure]{labelformat=empty} 
		\centering
		\subfloat[]{{\includegraphics[width=0.49\textwidth]{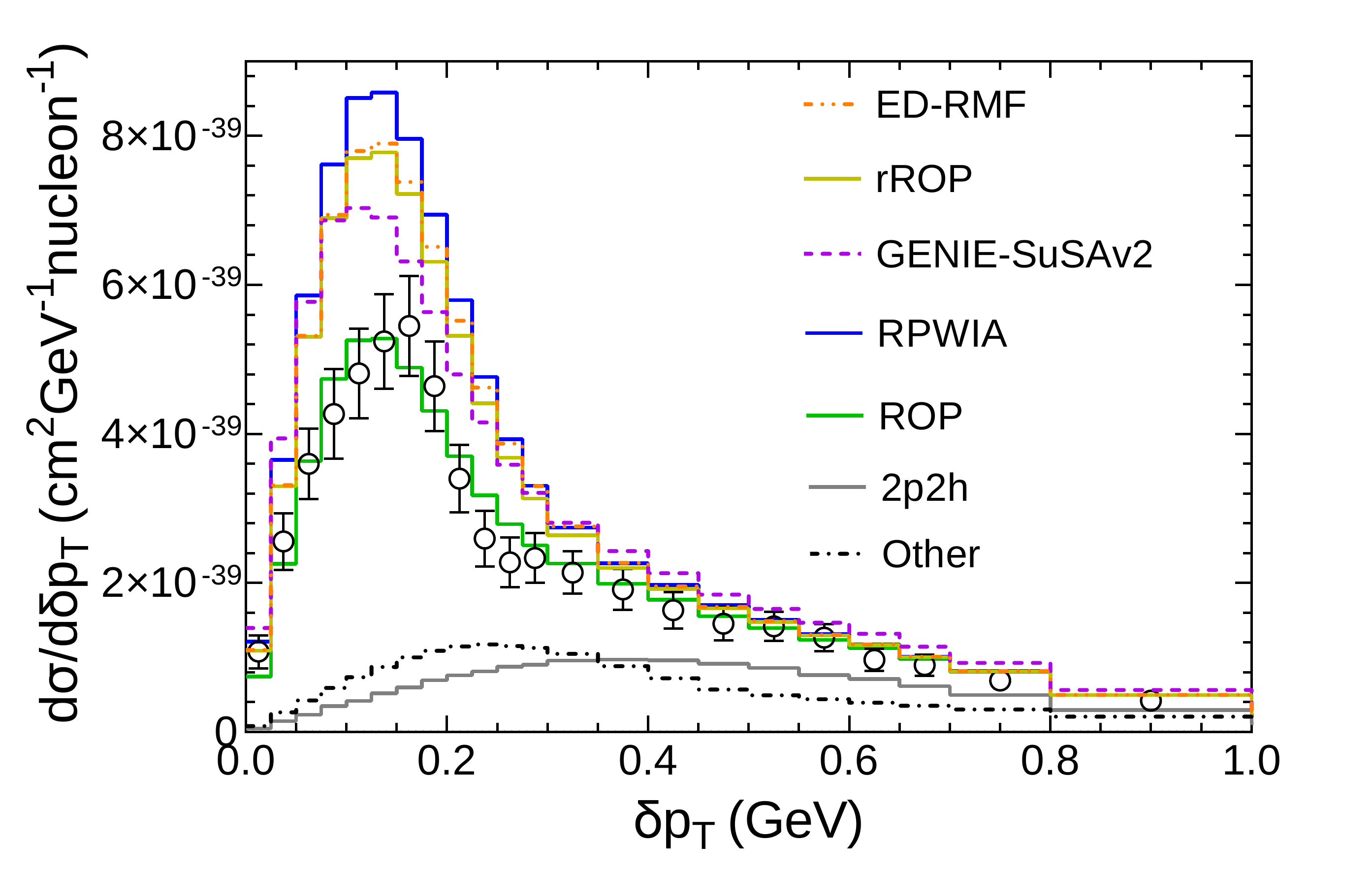}}}%
		\:
		\subfloat[]{{\includegraphics[width=0.49\textwidth]{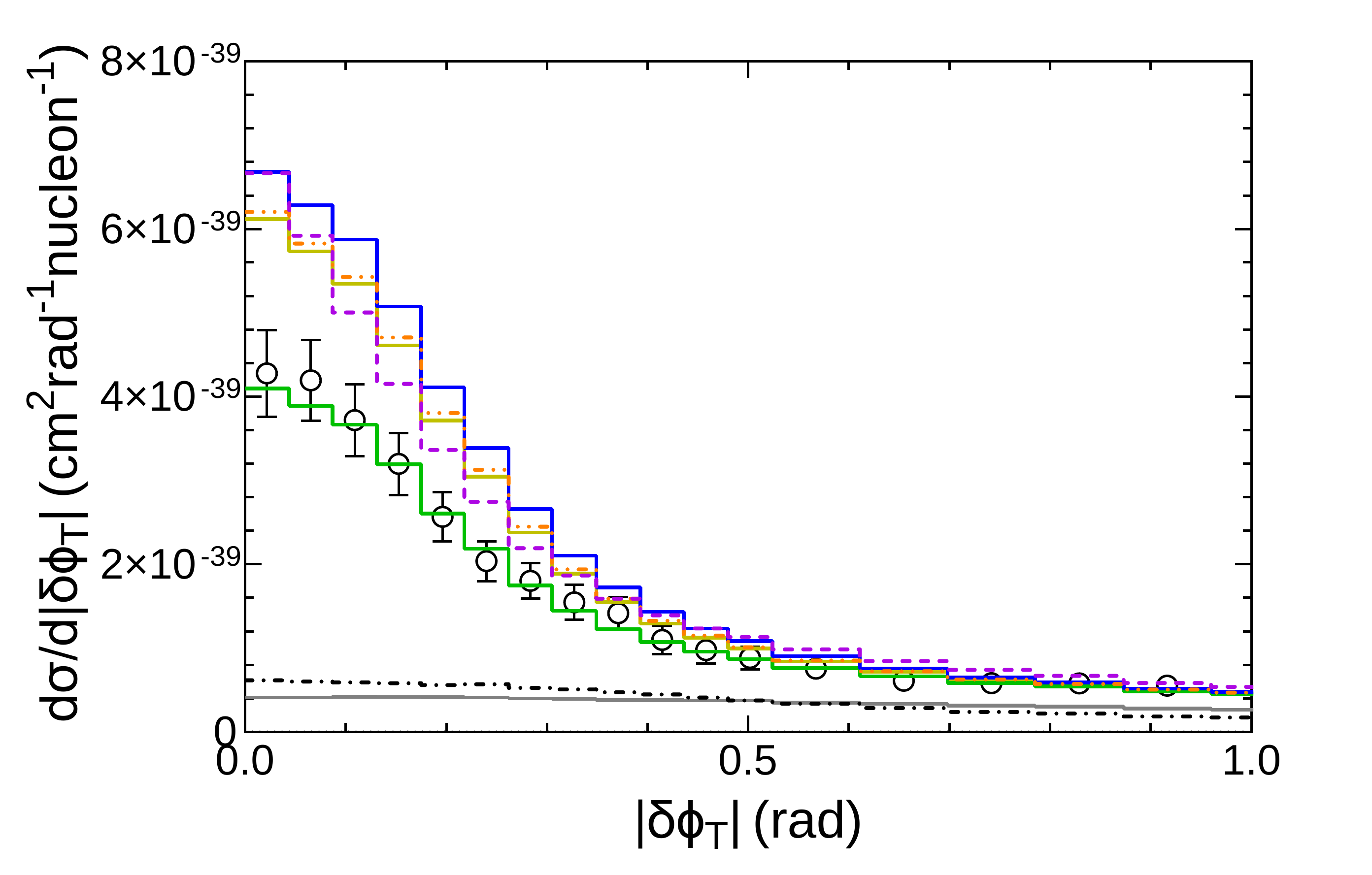}} }%
		\:
		\subfloat[]{{\includegraphics[width=0.49\textwidth]{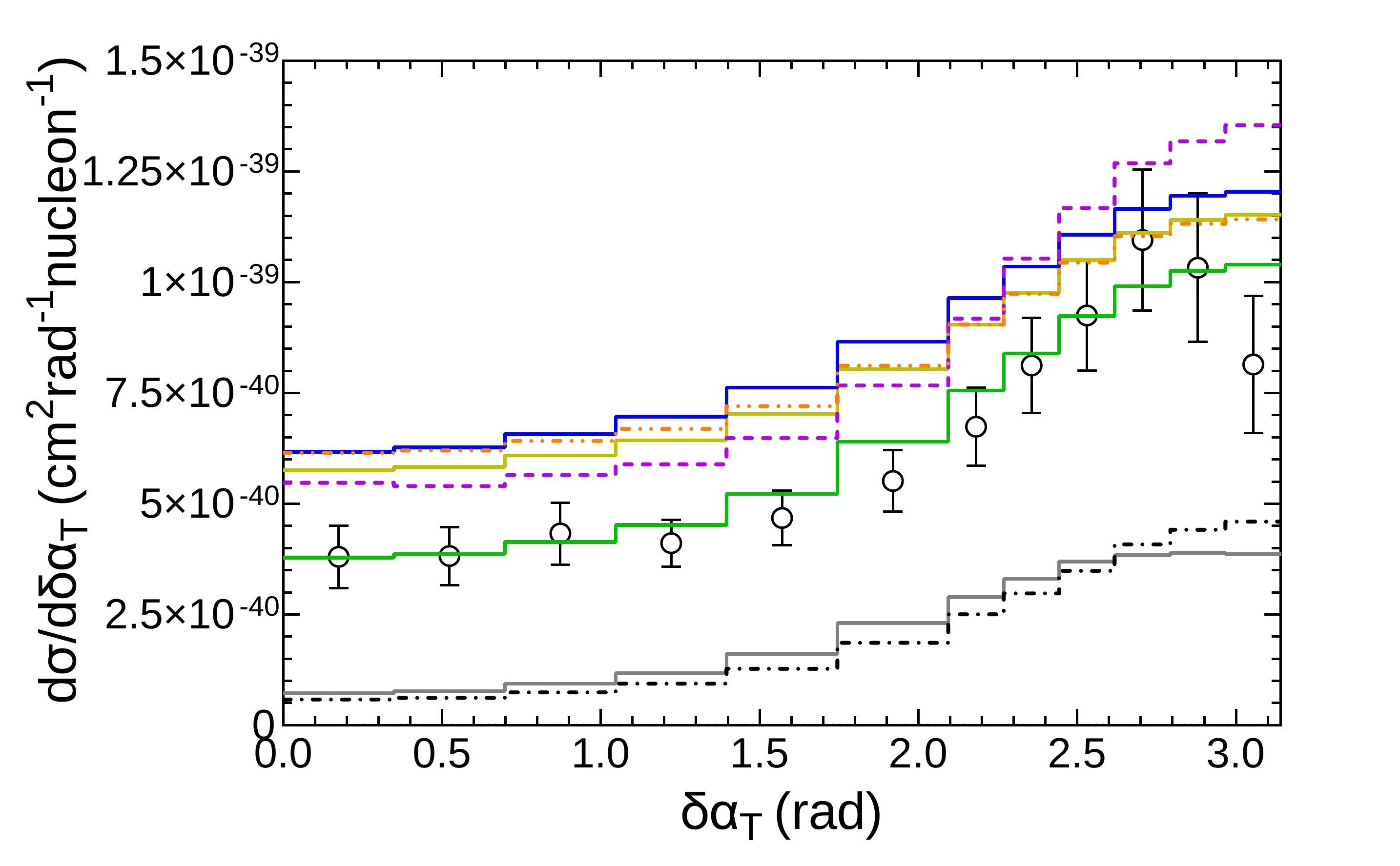}}}%
		\caption{\label{fig:STV minerva} MINER$\nu$A CC0$\pi$ semi-inclusive $\nu_\mu-^{12}$C cross sections as function of the transverse kinematic imbalances $\delta p_T$, $\delta\alpha_T$ and $\left|\delta\phi_T\right|$ defined in Eqs.~(\ref{tki:dpt}-\ref{tki:dphit}). All curves include the 2p2h and pion absorption contributions (also shown separately), evaluated using GENIE. The original paper from MINER$\nu$A was~\cite{PhysRevLett.121.022504} but the cross-section measurements shown here were taken from~\cite{PhysRevD.101.092001} which corrected a mismodelling in GENIE's elastic FSI that affected the cross-section measurements presented in the first paper.}
	\end{figure}

	In Fig.~\ref{fig:STV minerva} we compare the results of the different FSI models with the MINER$\nu$A cross-section measurements of the TKI distributions defined in Eqs.~(\ref{tki:dpt}-\ref{tki:dphit}). Even without adding the non-quasielastic contributions, all the models except ROP overestimate the data in the peak of the $\delta p_T$ distribution. In the high-momentum imbalance tail the contribution from the non-quasielastic channels is sufficient and clearly necessary to match any prediction with the experimental cross section. A similar situation is found for the $\delta\phi_T$ cross section where all the models except ROP overestimate the cross-section measurements near zero imbalance and a non-quasielastic contribution is clearly required to describe the tail of the distribution. 
	
    Regarding the $\delta\alpha_T$ results it is interesting to point out the appearance of a clear peak at large values in the MINER$\nu$A cross-section measurements that is not present in the T2K cross-section measurements shown in Fig.~\ref{fig:T2K semi-inclusive STV}, which might be caused by additional non-quasielastic contributions present in MINER$\nu$A due to the higher energy of the neutrinos. In case of T2K results shown in Fig.~\ref{fig:T2K semi-inclusive STV}, it has been shown~\cite{Dolan_2022} that the restriction of the proton momenta to be above $0.45$ GeV removes most of the interactions in which FSI plays an important role eliminating the peak at large $\delta\alpha_T$. The GENIE-SuSAv2 prediction and all the microscopic results except the ROP overestimate the cross-section measurement, although the shape of the rise in $\delta\alpha_T$ seems to be well described by the combination of FSI and non-quasielastic contributions.

\begin{figure}[!htbp]
		\captionsetup[subfigure]{labelformat=empty} 
		\centering
		{
			\subfloat[]{{\includegraphics[width=0.49\textwidth]{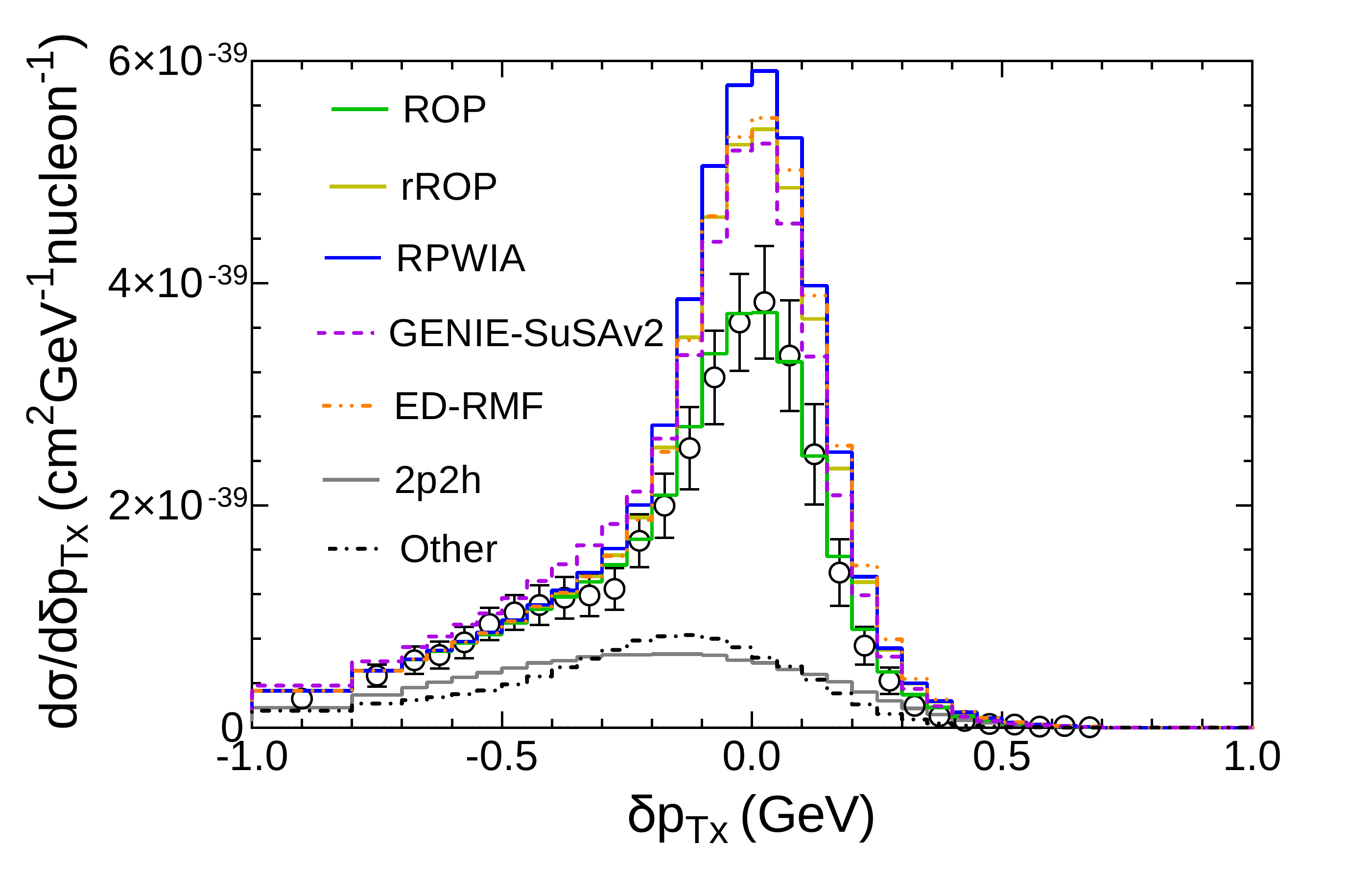}}}%
			\:
			\subfloat[]{{\includegraphics[width=0.49\textwidth]{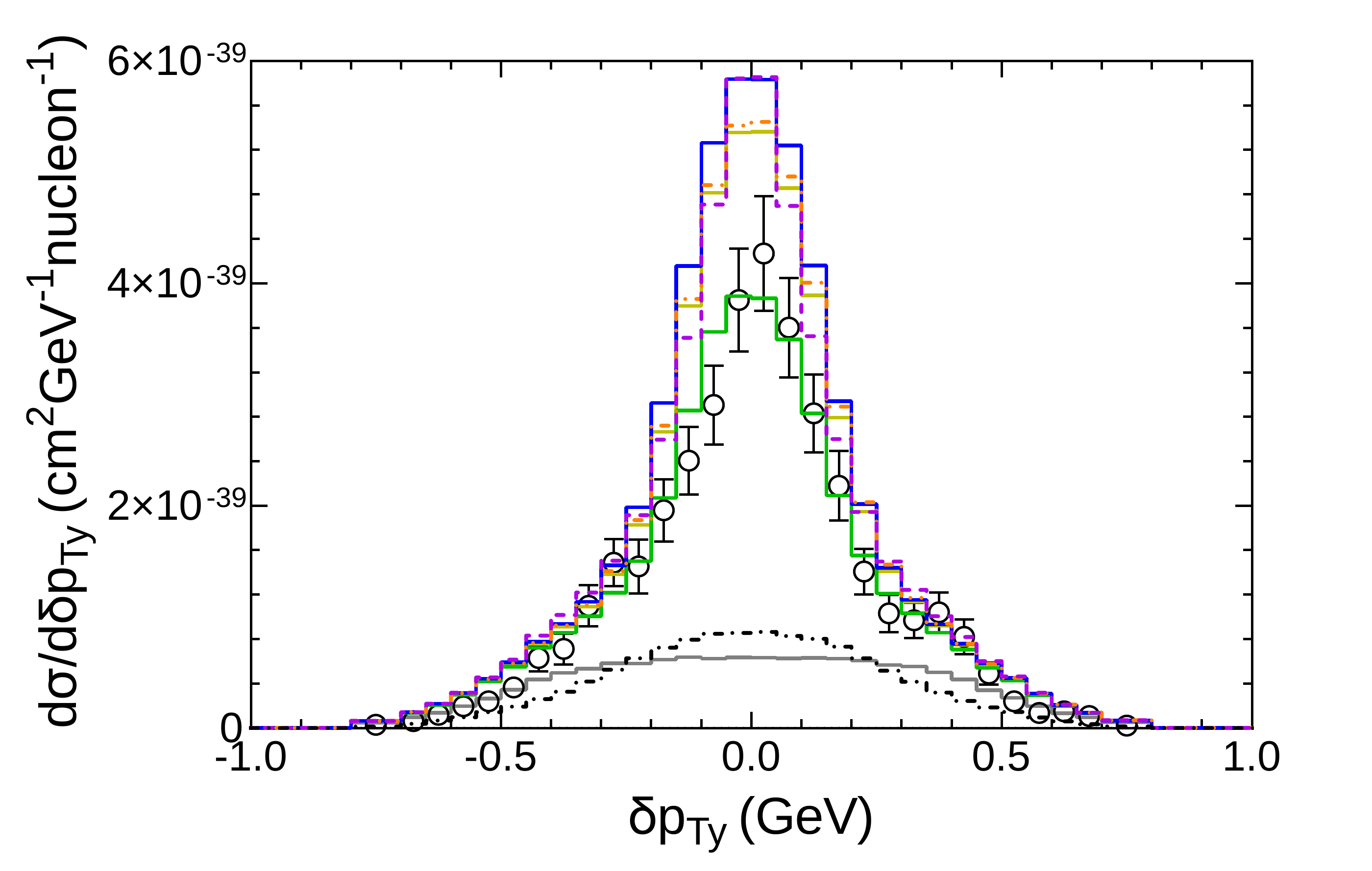}}}%
		}
		\caption{\label{fig:STV minerva projected} MINER$\nu$A CC0$\pi$ semi-inclusive $\nu_\mu-^{12}$C cross sections as function of the projections of $\delta p_T$ in the $x$ axis (parallel to the momentum transferred in the transverse plane) and in the $y$ axis (perpendicular to the momentum transferred in the transverse plane). All curves include the 2p2h and pion absorption contributions (also shown separately), evaluated using GENIE. Data taken from~\cite{PhysRevD.101.092001}. Notice that the convention used in~\cite{PhysRevD.101.092001} to define the $x$ and $y$ axis is the opposite to the convention used in this paper, which is the same adopted in~\cite{PhysRevD.104.073008}.}
	\end{figure}

	Additional projections on the plane perpendicular to the neutrino direction of the momentum imbalance $\delta p_T$ are presented in Fig.~\ref{fig:STV minerva projected}. If the interaction occurred on a free nucleon at rest, then we would expect a delta-function at $\delta p_T = 0$ because the muon and the proton in the final state would be perfectly balanced in that case. Therefore, the width of the QE peak is mostly consequence of the Fermi motion. In the PWIA $\delta {p_T}_y$ is exactly the projection on the y-axis of the initial nucleon momentum and, due to the isotropy of the nucleon momentum distribution, the $\delta {p_T}_y$ distribution is symmetrical. On the other hand $\delta {p_T}_x$ is parallel to the transferred momentum in the transverse plane, which is translated into a small shift of the peaks towards positive values of $\delta {p_T}_x$. In the $\delta {p_T}_x$ distribution the GENIE-SuSAv2 prediction is very similar to the ED-RMF and rROP models in the region of the peak, with all the results overestimating the cross-section measurements in this region except for ROP. Furthermore, GENIE-SuSAv2 differs slightly from the other models in the prediction of the asymmetric tail of the distribution where the non-quasielastic channels contribute more than the 1p1h channel. The non-quasielastic contributions seems to perfectly match the tails of the distribution when added to ROP, although alterations to the uncertain contributions coming from pion absorption (and, to some extent, 2p2h) may allow all the models other than RPWIA to match the cross-section measurements. Note that in all comparisons there is no evidence for a need for a significant enhancement of the 2p2h contribution, as is often suggested to be required by MINER$\nu$A measurements~\cite{MINERvA:2015ydy,PhysRevLett.121.022504}. The $\chi^2$ comparison for each TKI variable summarized in Table~\ref{table: chi2 minerva} shows that the best agreement is systematically obtained with the ROP model, although the $\chi^2/d.o.f$ values are far above 1 for all the TKI variables.

\section{\label{sec:5}Conclusions}

	We have compared the treatment of semi-inclusive neutrino-nucleus reactions within an unfactorized relativistic approach with the GENIE implementation of the SuSAv2-MEC model and latest T2K and MINER$\nu$A cross-section measurements. We have considered the one proton knockout process from a $^{12}$C target after the interaction with the $\nu_\mu$ beam used by T2K and MINER$\nu$A collaborations. For the microscopic method, the fully exclusive cross section as function of the final muon and proton is obtained without relying on the commonly used factorization approaches. The initial state is described by a RMF model with a realistic missing energy density that takes into account the depletion of the occupation numbers of the shells and correlations between bound nucleons, and different ways of including the distortion of the ejected nucleon by the residual nucleus, {\it i.e.} FSI, are considered, all of them fully consistent with special relativity by solving the Dirac equation with different potentials. 
	
	The implementation in GENIE of the SuSAv2-MEC model, which is based on RMF results and proved to describe successfully both  electron and neutrino inclusive reactions, can be used to make semi-inclusive predictions by assuming that the generation of the hadronic state is factorized from the rest of the process, although this is partially mitigated by a dependence of the removal energy with the momentum transferred. Since SuSAv2-MEC is a model integrated over all the nucleon variables, the initial state nucleon momentum is randomly sampled from a Fermi gas nuclear model and the propagation of the ejected nucleon through the residual nucleus is modeled using a semi-classical cascade model. To allow a more complete comparison with the cross-section measurements, the 2p2h predictions from the SuSAv2 model were added on top of all 1p1h results. In addition, the pion absorption contribution required for a complete comparison to experimental CC0$\pi$ measurements was also calculated from GENIE and added to the model predictions.
	
	Although it is difficult to draw precise conclusions about the individual effects of the approximations used to obtain semi-inclusive results with the SuSAv2 model, the microscopic calculation using a modified RMF potential (ED-RMF) improves the agreement with the experimental results at forward angles, {\it i.e.} low momentum and energy transfer, where scaling violations and low-energy effects not included in the SuSAv2 model are relevant. The measurement of correlations between the final muon and proton can help to explore the size and shape of the 2p2h contribution, for instance the comparison with the variables $\left|\Delta\mathbf{p}\right|$ and $\delta p_T$ in Figs.~\ref{fig:IV deltamodp}, \ref{fig:T2K semi-inclusive STV} and \ref{fig:STV minerva} where the 2p2h contribution is localized in the high imbalance tail. The $\left|\Delta\mathbf{p}\right|$ distribution is especially interesting for forward muon scattering angles because of the clear differences between the high and low muon momentum panels, with the former seemingly enhancing the relative 2p2h contribution over the latter. It is important to point out the limitations in the predictive power for nucleon kinematics of the implementation of the SuSAv2 2p2h model within GENIE, as well as of the pion absorption contribution. As shown in~\cite{Sobczyk_2020}, different models of the 2p2h channel can yield very different semi-inclusive cross-section predictions for this channel.
	
	The impact of the factorization approximation and the inconsistencies introduced by using different nuclear models in GENIE could be studied following the approach adopted in~\cite{nikolakopoulos2022benchmarking} to test the cascade model of NEUT against ROP results, where unfactorized rROP events generated with an initial state described by a RMF were fed to the neutrino generator and compared with ROP predictions under certain kinematic cuts. This could be the first step to implement in GENIE more exclusive models and more sophisticated descriptions of the initial state that go beyond a simple Fermi gas, which would improve the treatment of semi-inclusive reactions by neutrino event generators. An alternative is the strategy adopted by GiBUU~\cite{BUSS20121} and, more recently, by ACHILLES~\cite{https://doi.org/10.48550/arxiv.2205.06378}, which consists in using trajectories for the propagating nucleons that are dictated by the real part of an optical potential instead of straight lines.

\begin{acknowledgments}
		
	This work is part of the I+D+i projects PID2020-114687GB-100, and RTI2018-098868-B-I00 funded by MCIN (AED), and by the Junta de Andalucia (grants FQM160, SOMM17/61015/UGR and P20-01247); it is supported in part by the University of Tokyo ICRR's Inter-University Resarch Program FY2021 \& FY2022 (JAC, MBB, JMFP, GDM, RGJ, JMU), by the European Union's Horizon 2020 research and innovation programme under the Marie Sklodowska-Curie grant agreement No. 839481 (GDM), by the Project University of Turin (project BARM-RILO-21) and by INFN (national project NUCSYS) (MBB and JMFP), and by the government of Madrid and Complutense University under project PR65/19-22430 (R.G.-J). J.M.F.P. acknowledges support from a fellowship from the Ministerio de Ciencia, Innovación y Universidades. Program FPI (Spain).

\end{acknowledgments} 

\appendix
\section{\label{appendix}$\chi^2$ analysis}

\begin{figure*}[t]
		\centering
        \includegraphics[width=\textwidth,height=0.65\paperheight]{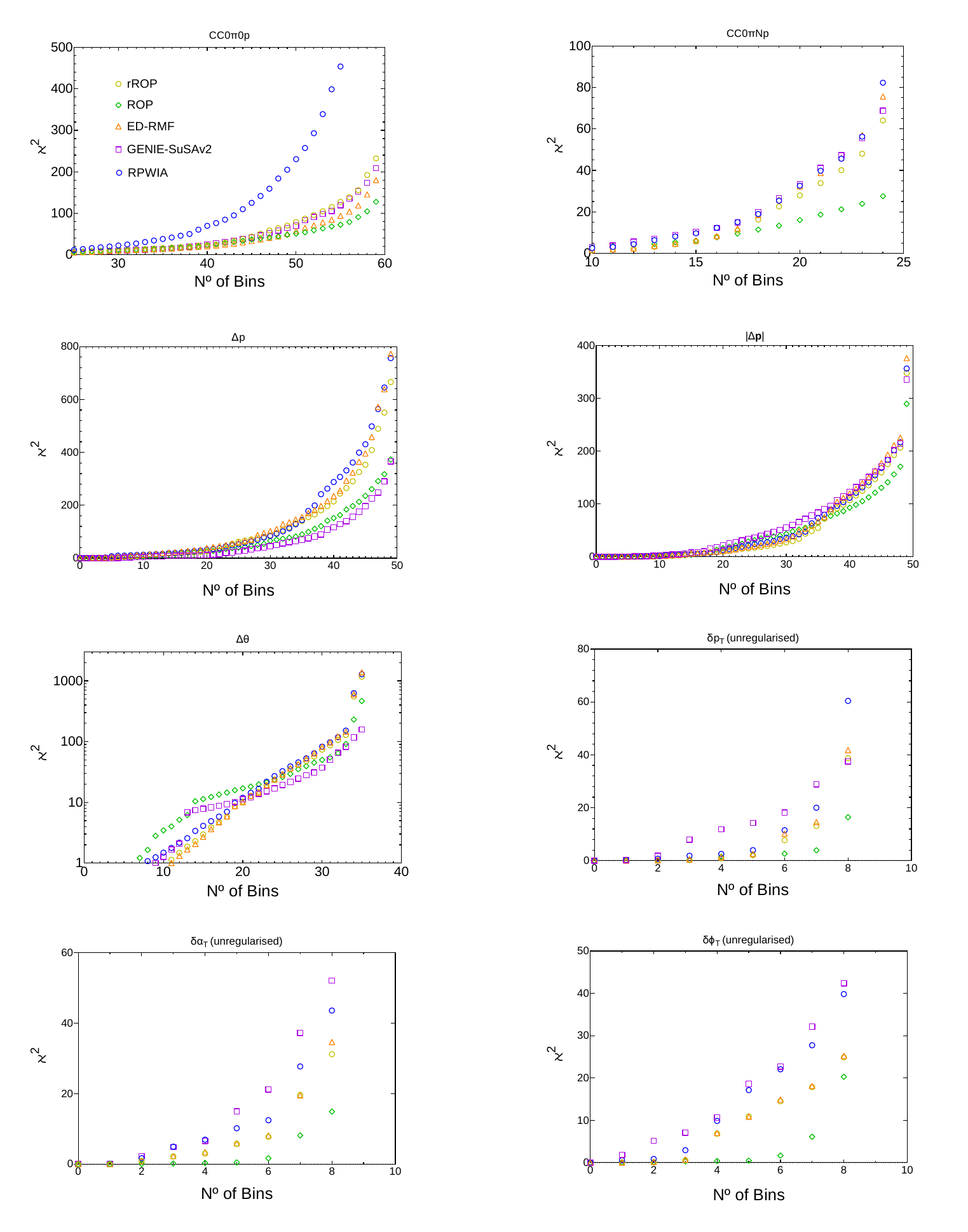}
		\caption{\label{fig:chi2} The evolution of the $\chi^2$ between each model and the T2K cross-section measurements when the bins contributing the largest $\chi^2$ are progressively removed. For example, the right-most points of the top left plot correspond to the total $\chi^2$ in Table~\ref{table: chi2 t2k} (calculated using all 59 bins reported by T2K), whilst the points just to the left of them show the $\chi^2$ once the bin contributing the largest $\chi^2$ is removed. Points further to the left remove more bins following the same rule. The bin contributing the largest $\chi^2$ is identified by re-calculating the $\chi^2$ after removing each bin (and its corresponding rows and columns in the covariance matrix) and choosing the largest.}
	\end{figure*}

Using the covariance matrices provided by the T2K~\cite{PhysRevD.98.032003} and MINER$\nu$A~\cite{PhysRevD.101.092001,PhysRevLett.121.022504} collaborations, we compute the $\chi^2$ between the predictions of the different models considered in this work and the cross section measurements. Within this section we use the unregularised TKI measurements provided by T2K~\cite{PhysRevD.98.032003}, as these are more suitable for quantitative $\chi^2$ analysis. Nevertheless, it was verified that differences in the $\chi^2$ analysis when using the regularised results are marginal. The $\chi^2$ of each model prediction for each measurement from T2K and MINERvA are compiled in Table~\ref{table: chi2 t2k} and Table~\ref{table: chi2 minerva}, respectively. The large $\chi^2$ with respect to the number of degrees of freedom for all measurements indicates a poor overall agreement, with the exception of ROP's good description of T2K's CC0$\pi$Np measurement and MINERvA's muon momentum measurement. However, it can be that the low $\chi^2$ is driven by poor agreement in a handful of outlying bins (for example the high muon momentum overflow bins reported by the T2K collaboration in the CC0$\pi0p$ results or the muon kinematic slices with very small cross section in the IV results). To study this, Fig.~\ref{fig:chi2} shows the evolution of the $\chi^2$ between each model and the T2K cross-section measurements when the bins contributing the largest $\chi^2$ are progressively removed. The bin contributing the largest $\chi^2$ is identified by re-calculating the $\chi^2$ after removing each bin (and its corresponding rows and columns in the covariance matrix) and choosing the largest one. The bins removed for each model are therefore different. 

Fig.~\ref{fig:chi2} immediately shows that the extreme $\chi^2$-values of all the microscopic models for the $\Delta\theta$ variable are mainly caused by two bins, which indeed are associated to very small values of the cross section (the -360 $ < \Delta\theta <$ -5 bin in the first two panels of Fig.~\ref{fig:IV deltatheta} which are not entirely shown to improve the readability of the plot). It also shows that the ROP model are in good agreement with the T2K CC0$\pi0p$ and TKI results once a few the worst bins are removed, many of which correspond to bins with extreme muon kinematics in the CC0$\pi0p$ case. Beyond some of the extreme muon kinematic bins, it is interesting to note that large RPWIA $\chi^2$-values are driven mostly by the forward going muon bins, which is not surprising given the strong suppressive effect of FSI in the corresponding low momentum transfer region.

For all T2K measurements other than the inferred variables it can be observed that the preference for ROP remains even after removing a significant fraction of bins. However, as discussed in Sec.~\ref{sec:4}, this may be dependent on the modelling of the non-quasielastic component of the interaction, which is currently subject to many approximations as described in~\ref{sec:3}.  

%	For the CC0$\pi1p$ and $\delta\phi_T$ results, Fig.~\ref{fig:chi2} further shows that the preference for ROP is not driven by only outlying bins. Conversely, the ROP preference is reduced for the other TKI by eliminating the lowest $\delta\alpha_T$ bin or the third $\delta p_T$ bin. 
	
	\vspace{0.3cm}
	
	\begin{table}[!htbp]
		\centering
		\begin{minipage}{0.5\textwidth}
			\begin{tabular}{cccccc} \toprule\toprule
				& rROP & ROP & RPWIA & ED-RMF & GENIE-SuSAv2  \\\midrule
				CC0$\pi$0p (59) & 232 & 127 &	1172 & 180 & 209\\ \midrule
				CC0$\pi$Np (24) & 64 & 28 & 82& 76& 69\\ \midrule
				$\Delta p$ (49)  & 666 & 373 & 756 & 773 & 366\\ \midrule
				$\Delta\theta$ (35) & 1170 & 466 & 1285 & 1379 & 159\\ \midrule
				$\Delta\theta^*$ (33) & 129 & 92 & 152 & 146 & 123\\ \midrule
				$\left|\Delta\mathbf{p}\right|$ (49)  & 348 & 290 & 357 & 376 & 336\\ \midrule
				$\delta p_T$ (8)  & 38 & 16 & 60 & 41 & 36\\ \midrule
				$\delta\alpha_T$ (8) & 29 & 13 & 41 & 33 & 49\\ \midrule
				$\delta\phi_T$ (8)  & 23 & 20 & 38 & 24 & 40\\
				\bottomrule\bottomrule
			\end{tabular}
		\end{minipage}\hfill
		\caption{$\chi^2$ values for different T2K topologies and variables. The degrees of freedom are given in brackets in the first column.  $\Delta\theta^*$ means that bins 0 and 5 were eliminated (-360 $ < \Delta\theta <$ -5 bin in the first two panels of Fig.~\ref{fig:IV deltatheta} which are not entirely shown to improve the readability of the plot).}
	\label{table: chi2 t2k}
	\end{table}

	\begin{table}[!htbp]
		\centering
		\begin{minipage}{0.5\textwidth}
			\begin{tabular}{cccccc} \toprule\toprule
				& rROP & ROP & RPWIA & ED-RMF & GENIE-SuSAv2  \\\midrule
				k' (32) & 61 & 30 & 79 & 64 & 71\\ \midrule
				$\theta_l$ (19) & 50 & 35 & 61 & 52 & 50\\ \midrule
				$p_N$ (25) & 106 & 61 & 120 & 115 & 116\\ \midrule
				$\theta_N^L$ (26) & 74 & 40 & 93 & 76 & 62\\ \midrule
				$\delta p_T$ (24)  & 126 & 51 & 165 & 132 & 114\\ \midrule
				$\delta p_{Tx}$ (33) & 128 & 58 & 161 & 144 & 85 \\\midrule
				$\delta p_{Ty}$ (32) & 127 & 84 & 155 & 131 & 118 \\\midrule
				$\delta\alpha_T$ (12) & 47 & 19 & 60 & 53 & 37\\ \midrule
				$\delta\phi_T$ (23)  & 126 & 99 & 155 & 130 & 93\\
				\bottomrule\bottomrule
			\end{tabular}
		\end{minipage}
		\caption{Same as in Table~\ref{table: chi2 t2k} but for MINER$\nu$A.}
		\label{table: chi2 minerva}
	\end{table}
	
\newpage
\bibliography{references}

\end{document}